\documentclass[aps,prl,reprint,superscriptaddress]{revtex4-1}
\usepackage{graphicx,color}
\usepackage{float}
\usepackage{amsmath}
\usepackage{xspace}
\usepackage{hyperref}
\usepackage[T1]{fontenc}
\newcommand{\degree}{\ensuremath{^\circ}\xspace}
\newcommand{\CBSCl}{Cu$_3$Bi(SeO$_3$)$_2$O$_2$Cl\xspace}
\newcommand{\CBSBr}{Cu$_3$Bi(SeO$_3$)$_2$O$_2$Br\xspace}
\newcommand{\CBSI}{Cu$_3$Bi(SeO$_3$)$_2$O$_2$I\xspace}
\newcommand{\CBSX}{Cu$_3$Bi(SeO$_3$)$_2$O$_2$$X$\xspace}

\newcommand{\TwoPls}{$^{2+}$\xspace}
\newcommand{\TNeel}{$T_\mathrm{N}$\xspace}
\newcommand{\TStr}{$T_{\mathrm{S}}$\xspace}

\begin{document}


\title{Magnetic and dielectric order in the kagome-like francisite \CBSCl }


\author{E. Constable}
\email[]{evan.constable@neel.cnrs.fr}
\altaffiliation[]{current address: Institute of Solid State Physics, Vienna University of Technology, Wiedner Hautptstr. 8-10 A-1040 Vienna, Austria}
\affiliation{Institut N\'eel, CNRS and Universit\'{e} Grenoble Alpes, 38042 Grenoble, France}
\author{S. Raymond}
\affiliation{INAC, Laboratoire Mod\'{e}lisation et Exploration des Mat\'{e}riaux, CEA and Universit\'{e} Grenoble Alpes, 38054 Grenoble, France}

\author{S. Petit}
\affiliation{CEA, CNRS, Universit\'{e} Paris-Saclay ,CE-Saclay F-91191 Gif-sur-Yvette, France}

\author{E. Ressouche}

\author{F. Bourdarot}
\affiliation{INAC, Laboratoire Mod\'{e}lisation et Exploration des Mat\'{e}riaux, CEA and Universit\'{e} Grenoble Alpes, 38054 Grenoble, France}

\author{J. Debray}
\affiliation{Institut N\'eel, CNRS and Universit\'{e} Grenoble Alpes, 38042 Grenoble, France}

\author{M. Josse}
\affiliation{CNRS, Universit\'e de Bordeaux, ICMCB, UPR 9048, F-33600 Pessac, France}

\author{O. Fabelo}
\affiliation{Institut Laue Langevin, CS 20156, 38042 Grenoble, France}

\author{H. Berger}
\affiliation{Institute of Physics, Ecole Polytechnique F\'{e}d\'{e}ral de Lausanne (EPFL), CH-1015 Lausanne, Switzerland}

\author {S. deBrion}
\author {V. Simonet}
\affiliation{Institut N\'eel, CNRS and Universit\'{e} Grenoble Alpes, 38042 Grenoble, France}


\date{\today}

\begin{abstract}

We report a single-crystal neutron diffraction and inelastic neutron scattering study on the spin 1/2 cuprate \CBSCl, complemented by dielectric and electric polarization measurements. The study clarifies a number of open issues concerning this complex material, whose frustrated interactions on a kagome-like lattice, combined with Dzyaloshinskii-Moriya interactions, are expected to stabilize an exotic canted antiferromagnetic order. In particular, we determine the nature of the structural transition occurring at 115 K, the magnetic structure below 25 K resolved in the updated space group, and the microscopic ingredients at the origin of this magnetic arrangement. This was achieved by an analysis of the measured gapped spin waves, which signifies the need of an unexpected and significant anisotropic exchange beyond the proposed Dzyaloshinskii-Moriya interactions. Finally, we discuss the mutliferroic properties of this material with respect to the space group symmetries.

\end{abstract}


\maketitle

\section{Introduction \label{sec:intro}}

The spin 1/2 kagome lattice has remained an important concept in frustrated magnetism stimulating both theoretical and experimental interests \cite{Chubukov_1992,Sachdev_1992,Shores_2005,Helton_2007,Yan_2011}. The characteristic geometry of this lattice, built from corner sharing triangles, promotes a natural competition between antiferromagnetic nearest-neighbor interactions. This favors a highly degenerate ground state, which when accompanied by quantum fluctuations, are the major ingredients of a quantum spin liquid, an exotic quantum phase whose exact nature has been the subject of many investigations \cite{Balents_2010}. Beyond the minimal kagome antiferromagnet model, macroscopic degeneracies and other exotic behaviors can be achieved when including further-neighbor competing interactions, even with leading ferromagnetic ones. An interesting case arises for ferromagnetic first and second neighbor interactions that compete with an antiferromagnetic interaction across the hexagons in the kagome lattice. New non-coplanar spin configurations can be favored either as short-range correlations in a spin liquid phase or crystallizing in exotic magnetic orders. This model also predicts several quantum chiral spin liquids (breaking the time reversal symmetry) \cite{Bieri_2015,Fak_2012}. These competing ferromagnetic and antiferromagnetic interactions have also been shown to be at the origin of other properties, such as magnetically induced ferroelectricity in the multiferroic KCu$_3$As$_2$O$_7$(OD)$_3$ compound \cite{Nilsen_2014}. In addition to the presence of several degrees of freedom and competing interactions, the role of weak anisotropies is expected to generate interesting behaviors in such kagome materials.

\begin{figure}
\centering
\includegraphics{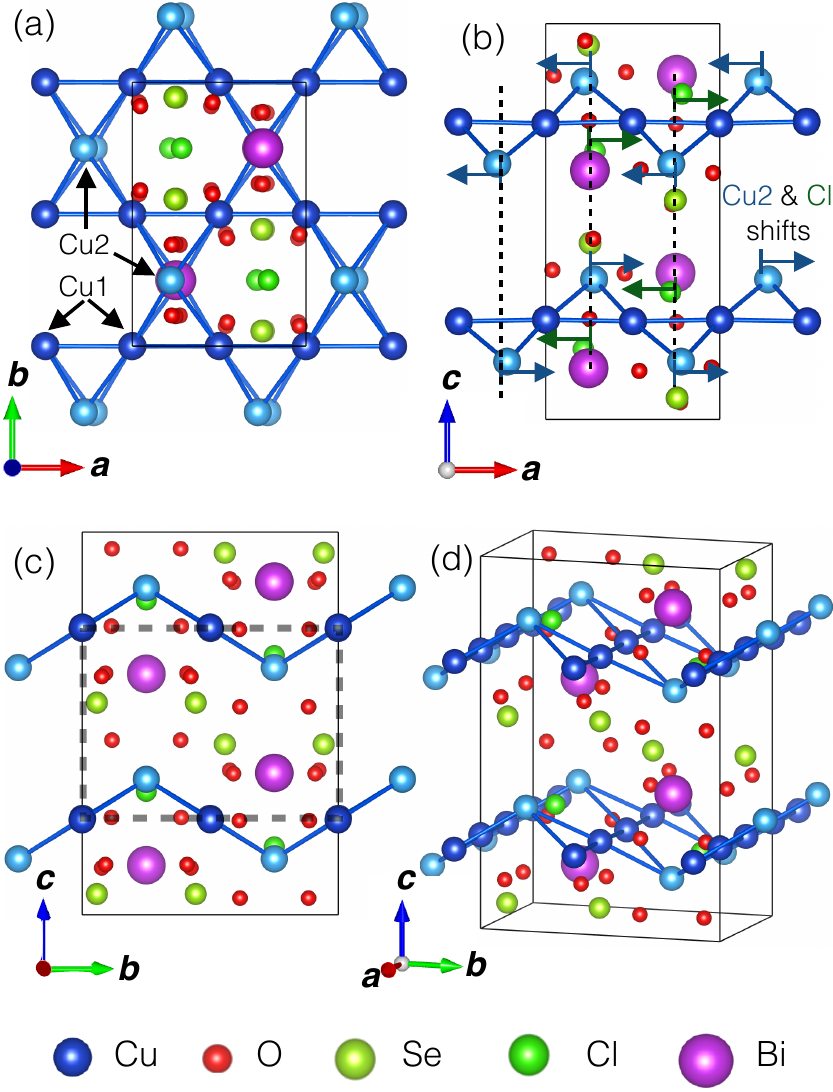}
\caption{$ab$ plane (a), $ac$ plane (b), $bc$ plane (c), and projected (d) views of the CBSCl crystal structure at 2 K in the $Pcmn$ space group with a doubling of the $c$ lattice parameter with respect to the room temperature $Pmmn$ structure. The dashed rectangle indicating the high temperature unit cell is shown in (c). The main consequences of the structural distortion are additional atomic shifts below 115 K of Cu2 and Cl. Their displacements are opposite in adjacent layers along the $c$ axis (a) and indicated by arrows in the $ac$ plane (b). The Cu\TwoPls pseudo-kagome layers are underlined in blue and their buckling is highlighted in the three dimensional depiction of the structure (d). Images were generated using VESTA software package \cite{Vesta_2011}.}
\label{fig:Nuc_crystal}
\end{figure}

In this context, a new class of materials attracting interest due to their unique frustrated magnetic state are the Francisites, namely, \CBSX, where $X$ is substituted by halogen ions Cl, Br or I. Originally discovered in natural Australian mineral deposits, the \CBSCl (CBSCl) compound was the first to be characterized \cite{Pring1990}. It was found to crystallize in the orthorhombic $Pmmn$ space group at room temperature. The structure features layers of CuO$_4$ bonded plackets that form a buckled kagome lattice stacked along the $c$ axis. The distorted kagome layers are separated by a network of SeO$_3$ triangular pyramids and Bi ions with the Cl atoms loosely confined inside the hexagonal voids of the kagome layers (Fig.~\ref{fig:Nuc_crystal}).

The magnetic properties are dictated by the presence of two distinct S = 1/2 Cu$^{2+}$ sites (Cu1 and Cu2) \cite{Millet2001} and by strong frustration due to competing ferro- and antiferromagnetic interactions on the kagome lattice \cite{Pregelj2012}. 
When considering only isotropic exchange and despite the dominant nearest-neighbor ferromagnetic interactions, a theoretical analysis has shown that this should lead to a classical degenerate ground state that survives quantum fluctuations \cite{Rousochatzakis2015}. This is due to the competition with the second-neighbor antiferromagnetic interaction present along the $b$ axis. However, the first experimental studies of CBSCl and closely related \CBSBr (CBSBr) have shown that these compounds indeed order antiferromagnetically below a N\'eel temperature \TNeel = 25 K \cite{Millet2001}.
It was thus argued that a tiny Dzyaloshinskii-Moriya (DM) interaction between the nearest-neighbors Cu1 and Cu2 spins is the stabilizing factor at the origin of a canted structure in the kagome layers, whereas a weak interlayer coupling produces the antiferromagnetic ordering along the $c$ axis.  As this latter coupling is very weak, the magnetic state is easily perturbed by an applied magnetic field along the $c$ axis. In this geometry,  a metamagnetic transition to a ferrimagnetic state occurs in a field of 0.7 T in which every second Cu\TwoPls spin layer rotates in alignment with the external field. Much higher fields ($\sim$8 T) are required to  reorient the magnetization along the magnetic field in the $ab$ plane \cite{Pregelj2012}.  A clear anisotropy is additionally observed between the $a$ and $b$ axes from magnetization measurements. These properties have been shown to be inherent in the francisite family featuring isostructural and magnetically equivalent CBSBr \cite{Pregelj2012} and Cu$_3$Y(SeO$_3$)$_2$O$_2$Cl \cite{Zakharov2014} compounds.

In CBSCl, the whole behavior is further complicated by the observation of phonon anomalies, absent in the Br and I francisites, suggesting that a structural distortion occurs at \TStr = 115 K, high above the magnetic ordering temperature \cite{Miller2012,Gnezdilov2016_arXiv}. It was further proposed that the low temperature structure might allow ferroelectricity, so that CBSCl could be multiferroic. Another pending issue is the discrepancy between the spin waves calculated using the Hamiltonian proposed by Rousochatzakis {\it et al.} \cite{Rousochatzakis2015}, combining isotropic exchange interactions and DM interactions, and the low-energy zone center modes measured by THz spectroscopy \cite{Miller2012}. Actually, experimental results are lacking to settle these issues since only the magnetic order of the Br compound has been determined by neutron diffraction and no spin wave dispersion has been measured in any members of the family.

Here, using single crystal neutron diffraction, we confirm the presence of a structural distortion at 115 K and we identify the low temperature structure in agreement with very recent X-ray diffraction measurements \cite{Prishchenko2017,Wu2017}.
Moreover, we refine the magnetic structure below 25 K in the correct space group, which is found to be rather similar to that of CBSBr. 
By single-crystal inelastic neutron scattering (INS), we investigate the spin wave spectrum in the ($a^*, c^*$) and ($b^*, c^*$) scattering planes, which reveals a global gap of the magnetic excitations. The comparison with spin wave calculations shows that, in addition to the weak Dzyaloshinskii-Moriya interaction required to stabilize the magnetic structure, a significant anisotropic exchange term is needed to account for this energy gap.
Finally, we have also measured the temperature dependence of the dielectric permittivity and of the electrical polarization along the $c$ axis and observed a behavior agreeing globally with the onset of an antiferroelectric phase in the vicinity of the structural transition.

\section{Experimental details\label{sec:methds}}

All neutron scattering experiments were performed on single crystals of CBSCl francisite grown by the standard chemical vapor-phase technique using the method described in reference \cite{Miller2012}.
The layered crystal structure favors growth along the [010] direction forming thin plate-like crystals in the $ab$ plane, with the $c$ axis normal to the sample faces.
The structural distortion was first investigated on the neutron Laue single-crystal diffractometer CYCLOPS at the Institut Laue Langevin (ILL) neutron facility on a 3.8$\times$2.3$\times$0.1 mm$^3$ single crystal. Then, neutron diffraction was performed on the CEA-CRG D23 thermal-neutron two-axis diffractometer at the ILL on a 6.7$\times$4.8$\times$0.7 mm$^3$ single crystal. The  diffractometer was operated in the 4-circle mode with the sample mounted in a 2 K displex refrigerator and using an incident and final wavelength of 1.173 \AA. Over 2603 reflections at temperatures of 150 K, 50 K and 2 K were collected for the nuclear and magnetic refinements.

Inelastic neutron scattering was performed on the 3 co-aligned crystals of dimensions 6.7$\times$4.8$\times$0.7 mm$^3$, 7.5$\times$3.8$\times$0.2 mm$^3$ and 5.1$\times$3.5$\times$0.3 mm$^3$ on the JCNS-CRG IN12, and ThALES cold-neutron triple axis spectrometers at the ILL, on the CEA-CRG IN22 thermal-neutron triple axis spectrometer at the ILL, and on the 2T-1 thermal-neutron triple axis spectrometer at the Laboratoire L\'eon Brillouin (LLB) neutron facility. The measurements in the ($a^*$, $c^*$) scattering plane were performed using the IN12, IN22 and 2T-1 spectrometers with $k_f=1.8$ \AA$^{-1}$, $k_f=2.662$ \AA$^{-1}$ and $k_f=2.662$ \AA$^{-1}$ respectively. The ($b^*$, $c^*$) scattering plane was investigated using the ThALES and IN22 spectrometers with $k_f=1.8$ \AA$^{-1}$ and $k_f=2.662$ \AA$^{-1}$ respectively. All these instruments were used in focusing geometry (no collimation) and higher order filters were installed (a velocity selector on IN12 and Thales, a graphite filter on IN22 and 2T).


Owing to the tabular nature of the samples, dielectric measurements performed at the Institut N\'eel (Grenoble) on a single-crystal (3.8$\times$2.3$\times$0.1 mm$^3$) were limited to the $c$ axis but additional measurements were performed on a powder sample compacted into a pellet. To determine the relative permittivity $\epsilon_r$, capacitive measurements were carried out by measuring the complex impedance of the sample with electrodes attached to its faces using a combination of silver paint and silver epoxy, and connected to an LCR meter (Agilent E4980A). The conditions chosen were an amplitude voltage of 1 V and a frequency of 20 kHz. The electric polarization $P$ was determined by integrating the pyroelectric current measured on the same sample using an electrometer (Keithley 6517A). To observe the effects of the structural distortion, the samples were  annealed at 250 K and cooled to 4.2 K at a rate of 3 K/min with a constant magnetic field and electric bias from $\pm$5 V to $\pm$200 V. At 4.2 K the electrical bias was removed and the temperature and magnetic field were maintained for 30 minutes to allow charge build-up on the electrodes to dissipate. The sample was then heated at 3 K/min in the same magnetic field up to 250 K while the current was measured as a function of time. Additional dielectric measurements were preformed at the ICMCB in Bordeaux. A HP4194 impedance bridge was used and the investigated frequency range was 1kHz to 1MHz. 
Measurements were performed in a PPMS (Quantum Design) for the sake of low temperature operation.

\section{Results \label{sec:results}}
\subsection{Structural transition}

The first experimental evidence to suggest the presence of a structural distortion in CBSCl was the observation of new phonon modes that appear in the far infrared spectra below \TStr = 115 K \cite{Miller2012}, a result that was further confirmed by Raman spectra \cite{Gnezdilov2016_arXiv}. 
This temperature also coincides with a deviation of the magnetic susceptibility from the high temperature Curie-Weiss law, indicating that this structural transition might be correlated with the onset of magnetic correlations \cite{Miller2012}.

\begin{figure}
\centering
\includegraphics{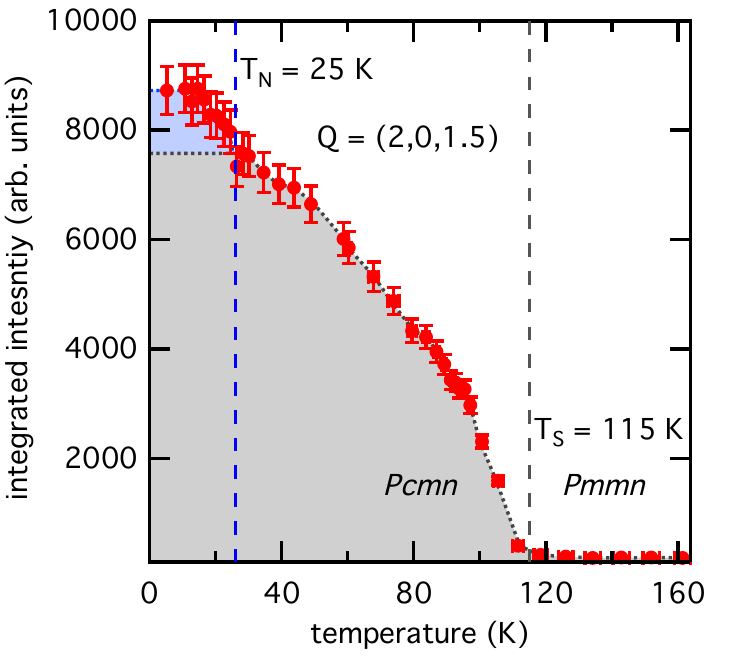}
\caption{Integrated neutron intensity of (2 0 1.5) Bragg peak, indexed in the room temperature space group, measured using the 2T-1 triple axis spectrometer in elastic mode. It rises with decreasing temperature below the structural transition and further increases below the magnetic ordering. The gray and blue shaded areas are guides for the eye and indicate the contributions from the structural distortion and magnetic phases respectively.}
\label{fig:T2_Bragg}
\end{figure}

 We first studied the structural distortion using Laue neutron diffraction on CYCLOPS that clearly showed a doubling of the $c$ lattice parameter below \TStr (not shown). Using the 2T-1 triple-axis spectrometer at zero energy, we then recorded $\omega$-scans (obtained by rotating the crystal) centered at the scattering vector {\bf Q} = (2 0 1.5) from 160 K to 5 K. Integration of the neutron counts as a function of temperature are shown in Fig.~\ref{fig:T2_Bragg}, which indeed reveals the emergence of a Bragg peak below 115 K and is consistent with a structural distortion involving a doubling along the $c$ axis as represented in Fig.~\ref{fig:Nuc_crystal} (c). Below 25 K, the peak gains further intensity from the magnetic ordering consistent with a ${\bf k} = (0, 0, 0)$ propagation vector in the new structure with a double $c$ lattice parameter (note that it would correspond to a ${\bf k} = (0, 0, 1/2)$ propagation vector as for CBSBr without the structural distortion).


To further investigate the nuclear structure below \TStr, we performed single-crystal neutron diffraction measurements on the D23 diffractometer, collecting Bragg reflections at 150, and 50 K with ${\bf k} = (0, 0, 0)$ in the orthorhombic cell with a $c$ lattice parameter doubled with respect to the room temperature structure. We carefully inspected the integrated intensities of the Bragg reflections and identified the following reflection conditions in the distorted phase.
\begin{equation*}
\begin{array}{lr}
h00: h = 2n& \quad hk0: h+k = 2n\\
0k0: k = 2n &\\
00l: l = 2n & \quad 0kl: l = 2n
\end{array}
\end{equation*}

\begin{figure}
\centering
\includegraphics{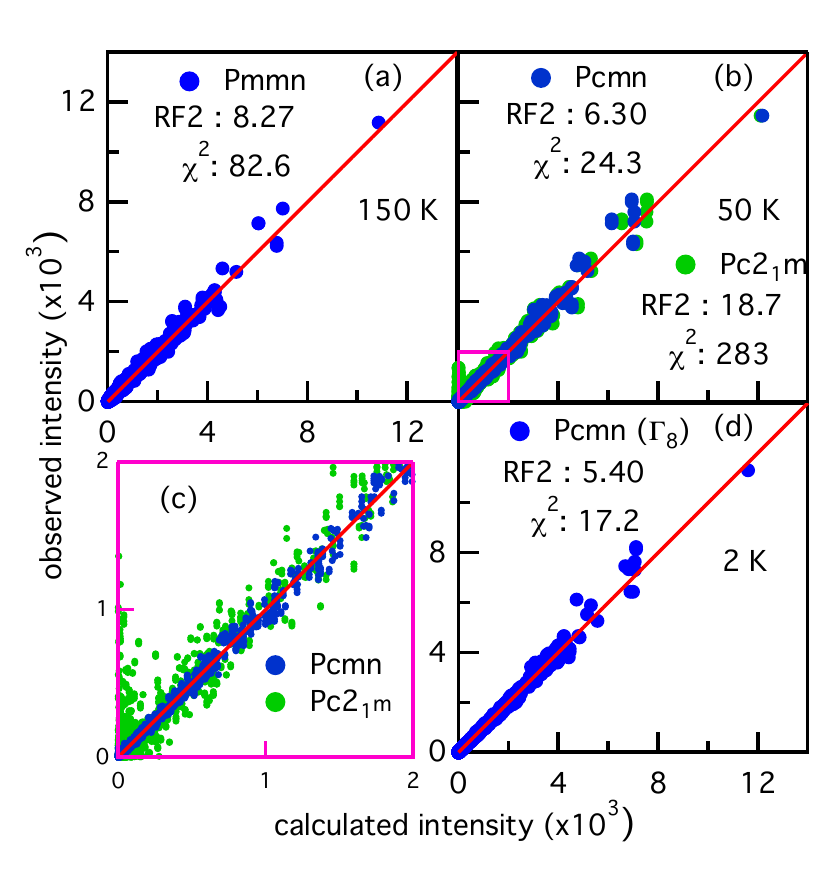}
\caption{Quality of the refinements of the nuclear structure at 150 K, above \TStr, in the $Pmmn$ space group (a), of the distorted nuclear structure at 50 K, below \TStr, in the $Pcmn$ and $Pc2_1m$ space groups (b-c), and of the superposed distorted nuclear structure and magnetic structure at 2 K, below \TNeel (d). These were refined in the $Pcmn$ space group and using the $\Gamma_8$ irreducible representation. The plots indicate the integrated intensity of the measured Bragg peaks versus the theoretically calculated values. A one-to-one dependence (red line) represents a perfect agreement. The agreement factors are reported on the plots. A zoom of the low intensity Bragg reflections for (b) is shown in (c), highlighting the discrepancy between the observed and the calculated intensities in particular for the new Bragg reflections produced by the distortion. 
}
\label{fig:FobsVsFcalc}
\end{figure}

Using the SUBGROUPS application of the Bilbao crystallographic server \cite{Aroyo_2006,Aroyo_2006_2,Bilbao_2011}, we identified 10 possible transformations from $Pmmn$ (No. 59) to an orthorhombic space group with a doubling of the $c$ lattice parameter. Of these possible space groups, only centrosymmetic $Pcmn$ (No. 62) and noncentrosymmetric $Pc2_1m$ (No. 33) fulfill the above conditions. Notably, the high temperature $Pmmn$ and previously proposed $P2_1mn$ space groups fail to meet the observed reflection conditions. 
We note that possible monoclinic and the triclinic non-centrosymmetric space groups  ($Pc$, $Pm$, $P2_1$, $P2$ and $P1$) can also be excluded based on the observed reflection conditions.
Using the data obtained on D23 at 50 K, we have attempted to fit the integrated intensities in both the $Pcmn$ and $Pc2_1m$ space groups using WinPLOTR of the FullProf software package \cite{Roisnel_2001}. The quality of the refinements is shown in Fig.~\ref{fig:FobsVsFcalc}. The best results are obtained for $Pcmn$ and indicate that displacements of the Cu2 and Cl atoms from $x/a$ = $\frac{1}{4}$ above \TStr to $x/a$ = $\frac{1}{4} +\Delta_{\mathrm{Cu2,Cl}}$ below \TStr, where $\Delta_{\mathrm{Cu2}}$=0.026 and $\Delta_{\mathrm{Cl}}$=0.031 at 50 K, are the defining distortions (see Fig.~\ref{fig:Nuc_crystal}(b)). The refined crystal structure parameters at 2 and 50 K are presented in Table~\ref{tab:Atms} along with the refined parameters at 150 K for the $Pmmn$ structure. 

\begin{table}
\setlength{\tabcolsep}{5pt}
\caption{Wyckoff positions, refined atomic positions, occupancy factors (calculated as the occupancy of the site divided by the maximum multiplicity) and isotropic displacement parameters in the low and high temperature phase of CBSCl. The results are presented for 2 K and 50 K in the $Pcmn$ (No.62) structure in the first and second rows of each atom respectively. The results at 150 K are for the $Pmmn$ (No.59) structure and shown in the third rows. The refined lattice parameters at 50 K are $a$ = 6.34080 \AA, $b$ = 9.62930 \AA{} and $c$ = 14.41840 \AA\ in the $Pcmn$ space group. The residuals of the refinement at 50 K are RF2 = 6.30 \% and $\chi^2$ = 24.3.}
\resizebox{\linewidth}{!}{\begin{tabular}{rcccccc}
\hline\hline
Atom & Site   & $x/a$         & $y/b$         & $z/c$      & Occ. & $B_{\mathrm{iso}}$\\
Bi$\,_{2\,\mathrm{K}}$   & $4c$   & 0.2465(1)     & $\frac{1}{4}$ & 0.12941(5) & 0.5  & 0.01(1) \\
   $_{50\,\mathrm{K}}$  &        & 0.2468(2)     & $\frac{1}{4}$ & 0.12941(6) & 0.5  & 0.03(2) \\
    $_{150\,\mathrm{K}}$ & $2a$   & $\frac{1}{4}$ & $\frac{1}{4}$ & 0.2406(3)  & 0.25 & 0.34(3) \\
Se$\,_{2\,\mathrm{K}}$   & $8d$   & 0.2447(1)     & 0.55707(7)    & 0.94570(4) & 1.0  & 0.08(1) \\
  $_{50\,\mathrm{K}}$   &        & 0.2453(1)     & 0.55701(8)    & 0.94567(4) & 1.0  & 0.11(1) \\
  $_{150\,\mathrm{K}}$   & $4e$   & $\frac{1}{4}$ & 0.5568(2)     & 0.6090(2)  & 0.5  & 0.47(3) \\
Cu1$\,_{2\,\mathrm{K}}$  & $8d$   & 0.0014(1)     & 0.0007(1)     & 0.25201(5) & 1.0  & 0.12(1) \\
  $_{50\,\mathrm{K}}$   &        & 0.0013(1)     & 0.0006(1)     & 0.25179(6) & 1.0  & 0.19(1) \\
  $_{150\,\mathrm{K}}$   & $4c$   & 0             & 0             & 0          & 0.5  & 0.62(3) \\
Cu2$\,_{2\,\mathrm{K}}$  & $4c$   & 0.2218(1)     & $\frac{1}{4}$ & 0.85401(5) & 0.5  & 0.11(1) \\
  $_{50\,\mathrm{K}}$   &        & 0.2242(2)     & $\frac{1}{4}$ & 0.85408(6) & 0.5  & 0.20(2) \\
  $_{150\,\mathrm{K}}$   & $2a$   & $\frac{1}{4}$ & $\frac{1}{4}$ & 0.7922(3)  & 0.25 & 0.73(4) \\
Cl$\,_{2\,\mathrm{K}}$   & $4c$   & 0.2167(1)     & $\frac{3}{4}$ & 0.17757(5) & 0.5  & 0.47(1) \\
  $_{50\,\mathrm{K}}$   &        & 0.2192(2)     & $\frac{3}{4}$ & 0.17735(6) & 0.5  & 0.68(2) \\
  $_{150\,\mathrm{K}}$   & $2b$   & $\frac{1}{4}$ & $\frac{3}{4}$ & 0.1474(3)  & 0.25 & 2.15(4) \\
O1$\,_{2\,\mathrm{K}}$   & $8d$   & 0.2480(1)     & 0.11359(9)    & 0.75409(5) & 1.0  & 0.15(1) \\
 $_{50\,\mathrm{K}}$    &        & 0.2484(2)     & 0.1136(1)     & 0.75409(6) & 1.0  & 0.30(1) \\
  $_{150\,\mathrm{K}}$   & $4e$   & $\frac{1}{4}$ & 0.1137(2)     & 0.9914(3)  & 0.5  & 0.37(3) \\
O2$\,_{2\,\mathrm{K}}$   & $8d$   & 0.0462(1)     & 0.5905(1)     & 0.86874(6) & 1.0  & 0.26(2) \\
  $_{50\,\mathrm{K}}$   &        & 0.0459(2)     & 0.5900(1)     & 0.86904(7) & 1.0  & 0.30(1) \\
  $_{150\,\mathrm{K}}$   & $8g$   & 0.0418(2)     & 0.5832(2)     & 0.7542(2)  & 1.0  & 0.85(3) \\
O2'$\,_{2\,\mathrm{K}}$  & $8d$   & 0.0378(1)     & 0.5761(1)     & 0.37739(6) & 1.0  & 0.26(2) \\
  $_{50\,\mathrm{K}}$   &        & 0.0382(2)     & 0.5766(1)     & 0.37709(7) & 1.0  & 0.30(1) \\
  $_{150\,\mathrm{K}}$   &  --    &    --         &    --         &   --       & --   & --      \\
O3$\,_{2\,\mathrm{K}}$   & $8d$   & 0.2268(1)     & 0.11608(9)    & 0.95645(5) & 1.0  & 0.32(1) \\
  $_{50\,\mathrm{K}}$   &        & 0.2285(2)     & 0.1162(1)     & 0.95635(6) & 1.0  & 0.30(1) \\
  $_{150\,\mathrm{K}}$   & $4e$   & $\frac{1}{4}$ & 0.1162(3)     & 0.5881(3)  & 0.5  & 1.07(4) \\
\hline\hline
\end{tabular}}
\label{tab:Atms}
\end{table}

Our results are in agreement with the recent X-ray refinement of the low temperature structure by Prishchenko {\it et al.} and discards the polar space group $P2_1mn$ proposed initially to account for the additional 16 phonon modes observed below \TStr \cite{Miller2012}. Actually, a doubling of the $c$ lattice parameter, that was not considered in this previous work, provides a natural explanation to the presence of the new phonon modes as an expansion of the unit cell will increase the number of normal modes it supports. Therefore a loss of centrosymmetry is not required to explain the phonon anomalies. 

As shown in the study of Prishchenko et al. \cite{Prishchenko2017}, and by the previous literature on francisites including neutron \cite{Pregelj2012} and Raman \cite{Gnezdilov2016_arXiv} scattering experiments, the structural transition observed in CBSCl is not found in the CBSBr and \CBSI (CBSI) forms of francisite. Differing only in the types of halogen atoms in the chemical formula, one can reason that it is the size of the halogen atoms in the francisites that is the source of the structural transition in CBSCl. In the francisites, the halogen atom's position at the center of the kagome hexagons, places it in a potential void where large atomic motion is possible. Unlike Br and I, which are larger atoms than Cl and therefore relatively stable in their positions, Cl will be more free to displace from its position. This, as well as the buckled geometry of the kagome hexagon that places the Cu2 ions out of the kagome plane, promotes an instability of both the Cl and Cu2 ions. Magnetic correlations across the kagome lattice as well as thermal lattice contraction put the Cl and Cu2 ions in energetically unfavorable positions within the $Pmmn$ structure. Therefore, a collective displacement of the Cu2 and Cl ions to a lower energy state occurs. This is consistent with our crystal refinement results, which show antiparallel shifts for Cu2 and Cl along the $a$ crystallographic direction, reducing the Cu2-Cl distance from 3.20 \AA\  at 150 K to 2.82 \AA\ at 50 K. The new positions are no longer described by the symmetries of the $Pmmn$ space group, rather falling under the $Pcmn$ description with a twice as large $c$ cell parameter. Including this, we also find that as a result of these displacements, small shifts in the O2, O2' and O3 ions occur. Importantly, this results in a distortion of the SeO$_3$ triangular pyramid. We note that, while no considerable distortions are observed below our refinement at 50 K, it is the oxygen atoms that have the biggest shifts in position when comparing results at 50 K and 2 K.

\subsection{Magnetic order}

\begin{table*}
\caption{Magnetic configurations associated to the 8 irreducible representations for the Cu1 and Cu2 $S = 1/2$ magnetic sites of the $Pcmn$ space group with $\mathbf{k} = (0,0,0)$ propagation vector calculated using BasIreps of the FullProf software package \cite{Rodriguez_Carvajal_1993}. The $u$, $v$, $w$, $u'$ and $v'$ parameters are refinable parameters for the magnetic moment components.}
\resizebox{\linewidth}{!}{
\label{tab:MagIrreps}
\begin{tabular}{c|cccccccc}
\hline \hline
 Site Cu1 & $\Gamma_1$ & $\Gamma_2$ & $\Gamma_3$ & $\Gamma_4$ & $\Gamma_5$ & $\Gamma_6$ & $\Gamma_7$ & $\Gamma_8$\\
\hline
$x,y,z$ & $u,v,w$ & $u,v,w$ & $u,v,w$ & $u,v,w$ & $u,v,w$ & $u,v,w$ & $u,v,w$ & $u,v,w$ \\

$-x+\frac{1}{2},-y+\frac{1}{2},z+\frac{1}{2}$ &  $-u,-v,w$ & $-u,-v,w$  & $-u,-v,w$  & $-u,-v,w$ & $u,v,-w$ & $u,v,-w$ & $u,v,-w$ & $u,v,-w$ \\

$x+\frac{1}{2},-y,-z+\frac{1}{2}$ &  $u,-v,-w$ & $u,-v,-w$ & $-u,v,w$ & $-u,v,w$ & $u,-v,-w$ & $u,-v,-w$ & $-u,v,w$ & $-u,v,w$ \\

$-x,y+\frac{1}{2},-z$ &  $-u,v,-w$ & $-u,v,-w$ & $u,-v,w$ & $u,-v,w$ & $u,-v,w$ & $u,-v,w$ & $-u,v,-w$ & $-u,v,-w$ \\

$-x,-y,-z$ &  $u,v,w$ & $-u,-v,-w$  & $u,v,w$ & $-u,-v,-w$ & $u,v,w$ & $-u,-v,-w$ & $u,v,w$ & $-u,-v,-w$ \\

$x+\frac{1}{2},y+\frac{1}{2},-z+\frac{1}{2}$ &  $-u,-v,w$ & $u,v,-w$ & $-u,-v,w$ & $u,v,-w$ & $u,v,-w$ & $-u,-v,w$ & $u,v,-w$ & $-u,-v,w$ \\

$-x+\frac{1}{2},y,z+\frac{1}{2}$ &  $u,-v,-w$ & $-u,v,w$ & $-u,v,w$ & $u,-v,-w$ & $u,-v,-w$ & $-u,v,w$ & $-u,v,w$ & $u,-v,-w$ \\

$x,-y+\frac{1}{2},z$ &  $-u,v,-w$ & $u,-v,w$ & $u,-v,w$ & $-u,v,-w$ & $u,-v,w$ & $-u,v,-w$ & $-u,v,-w$ & $u,-v,w$ \\
\hline
 Site Cu2 & $\Gamma_1$ & $\Gamma_2$ & $\Gamma_3$ & $\Gamma_4$ & $\Gamma_5$ & $\Gamma_6$ & $\Gamma_7$ & $\Gamma_8$\\
\hline
$x,y,z$ & $0,u',0$ & $u',0,v'$ & $u',0,v'$ & $0,u',0$ & $u',0,v'$ & $0,u',0$ & $0,u',0$ & $u',0,v'$ \\

$-x+\frac{1}{2},-y+\frac{1}{2},z+\frac{1}{2}$ & $0,-u',0$ & $-u',0,v'$  & $-u',0,v'$ & $0,-u',0$ & $u',0,-v'$ & $0,u',0$ & $0,u',0$ & $u',0,-v'$ \\

$x+\frac{1}{2},-y,-z+\frac{1}{2}$ & $0,-u',0$ & $u',0,-v'$  & $-u',0,v'$ & $0,u',0$ & $u',0,-v'$ & $0,-u',0$ & $0,u',0$ & $-u',0,v'$ \\

$-x,y+\frac{1}{2},-z$ & $0,u',0$ & $-u',0,-v'$  & $u',0,v'$ & $0,-u',0$ & $u',0,v'$ & $0,-u',0$ & $0,u',0$ & $-u',0,-v'$ \\
\hline \hline
\end{tabular}}
\end{table*}

To determine the magnetic structure, a further set of neutron Bragg reflections were collected and analyzed at 2 K, below the magnetic transition. While a previous single crystal neutron diffraction refinement of CBSBr has revealed a magnetic structure consistent with the macroscopic magnetic properties of both the Br and Cl compounds \cite{Pregelj2012}, no neutron diffraction studies have been reported so far on the Cl compound. One can wonder whether the structural distortion, absent in the Br compound, has an influence on the magnetic order of the Cl compound. In CBSCl the magnetic phase exists within the $Pcmn$ structure which is compatible with a larger set of irreducible representations (IR), with respect to the undistorted case of CBSBr. Through a second order phase transition and for coupled Cu1 and Cu2 spins, one of these IR should correspond to the observed magnetic structure. The 8 possible IRs, determined using BasIreps of the FullProf software package \cite{Rodriguez_Carvajal_1993}, are reported in Table~ \ref{tab:MagIrreps}.

We have tested all possibilities and obtained the best refinement for the magnetic structure described by the $\Gamma_8$ IR on both Cu1 and Cu2 sites, giving rise to the $Pcm'n$ magnetic space group ($Pnm'a$ in the standard setting). Contrary to what has been refined for the CBSBr compound, where canting along $a$ of the Cu2 sites is forbidden in the $Pmmn$ structure \cite{Pregelj2012}, the $\Gamma_8$ IR here allows for a magnetic component along the $a$ axis for both the Cu1 and Cu2 sites. These components are refined by the $u$ and $u'$ parameters in Table~\ref{tab:MagIrreps} respectively. In our refinement of CBSCl however, we were unable to experimentally distinguish between a fit where $u$ and $u'$ is varied or when they are fixed to $u=u'=0$. When $u$ and $u'$ are allowed to vary we do indeed refine a structure featuring a very small alternation of the Cu1 spins along $a$ at an angle of 3.8\degree and $ac$ plane canted Cu2 spins at an angle of 32.6\degree from $c$ towards $a$. However, such a magnetic structure with finite $u$ and $u'$ components disagrees with our mean field calculations and spin wave measurements presented later in the article which suggests no canting along the $a$ axis for either the Cu1 or Cu2 sites.
We therefore proceeded using a $\Gamma_8$ refined magnetic structure with $u=u'=0$ fixed for the Cu1 and Cu2 sites. The quality of the refinement is shown in Fig.~\ref{fig:FobsVsFcalc} (b). The refined magnetic components are then 
$m_y=0.76(3)\mu_{\mathrm{B}}$ and $m_z=-0.46(5)\mu_{\mathrm{B}}$ for Cu1 (total magnetic moment of 0.89(4)$\mu_{\mathrm{B}}$) and $m_z=m_{tot}=1.03(9)\mu_{\mathrm{B}}$ for Cu2. The magnetic structure is depicted in Fig~\ref{fig:Mag_Struct}. It is overall similar to the previously reported structure of CBSBr \cite{Pregelj2012}: the Cu2 spins are along the $c$ axis and the Cu1 spins are confined  to the $bc$ plane alternating along $b$ with a canting of 59\degree$\pm$4\degree from $c$ towards $b$. One particular difference is the increased canting of the Cu1 spins with 59\degree for CBSCl and 50\degree for CBSBr \cite{Pregelj2012}. This difference is possibly a results of the structural distortion in CBSCl, and is discussed later in the context of the magnetic Hamiltonian.

\subsection{Hamiltonian and magnetic excitations}

\begin{figure}
\centering
\includegraphics{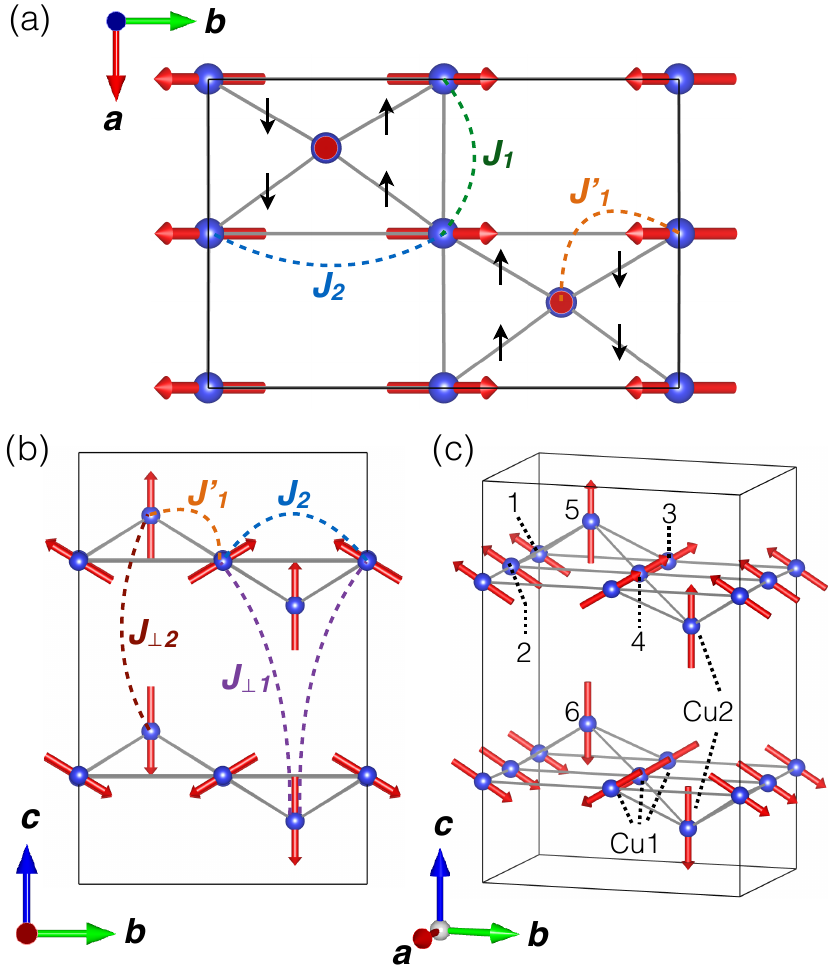}
\caption{Refined canted magnetic structure at 2 K using the $\Gamma_8$ irreducible representation with $u=u'=0$ of Table~\ref{tab:MagIrreps}. The Cu1 and Cu2 spins are shown in the $ab$ plane (a), in the $bc$ plane (b), and in a three dimensional representation (d). The spins are indexed according to equation~\ref{eq:Hamiltonian}. The exchange interactions in the plane and between the planes are reported ((a) and (b)), as well as the dominant component of the DM interaction between the nearest-neighbor Cu1 and Cu2 spins (black arrows) in panel (a).
Images were generated using VESTA software package \cite{Vesta_2011}.}
\label{fig:Mag_Struct}
\end{figure}

The originality of this magnetic structure, antiferromagnetic stacking of canted spins, was already addressed by Rousochatzakis \textit{et al.} in relation to the role of magnetic frustration \cite{Rousochatzakis2015}. Their first theoretical analysis was aimed at determining the principal magnetic interactions in the francisites using density functional theory. In their study it was found that the nearest neighbor interactions, $J_1$ between Cu1 spins along $a$ and $J_1'$ between Cu1 and Cu2 spins diagonally along $a$-$b$, are dominantly ferromagnetic. A comparably large next nearest-neighbor interaction between the Cu1 spins along $b$, $J_2$, is antiferromagnetic and provides the source of frustration. Weak interlayer antiferromagnetic interactions  $J_{\perp1}$ and $J_{\perp2}$ are also present between the Cu1 and Cu2 sites. The approximate paths of these primary interactions are indicated in Fig~\ref{fig:Mag_Struct} (a) and (b). To stabilize the magnetic structure, a DM interaction is introduced between the Cu1 and Cu2 spins with a dominant $a$ axis component that alternates along $b$, depicted by black arrows in Fig~\ref{fig:Mag_Struct} (a).
Longer range interactions between the Cu1 spins along $a$ and $b$ are determined negligible.
The resulting proposed magnetic Hamiltonian is then,
\begin{eqnarray}
\label{eq:Hamiltonian}
H_0 &=& J_1\sum_{\substack{i = 1,3\\ j = 2,4}} \mathbf{S_i\cdot S_j} + J_1'\sum_{\substack{i = 1-4\\j = 5 }}\mathbf{S_i\cdot S_j}
+ J_2\sum_{\substack{i = 1,2\\j = 3,4}} \mathbf{S_i \cdot S_j} \nonumber\\*
&+& J_{\perp1}\sum_{\substack{i = 1-4\\j = 6}} \mathbf{S_i \cdot S_j}
+ J_{\perp2}\sum_{\substack{i = 5\\j = 6}} \mathbf{S_i \cdot S_j} \nonumber\\*
&+& \mathbf{D}\ \cdot\sum_{\substack{i = 1-4\\j = 5}} \mathbf{S_i \times S_j} ,
\end{eqnarray}
with $\mathbf{D}$ the DM vector and the indexes corresponding to the labeled spins in Fig.~\ref{fig:Mag_Struct} (c).

\begin{figure*}
\centering
\includegraphics{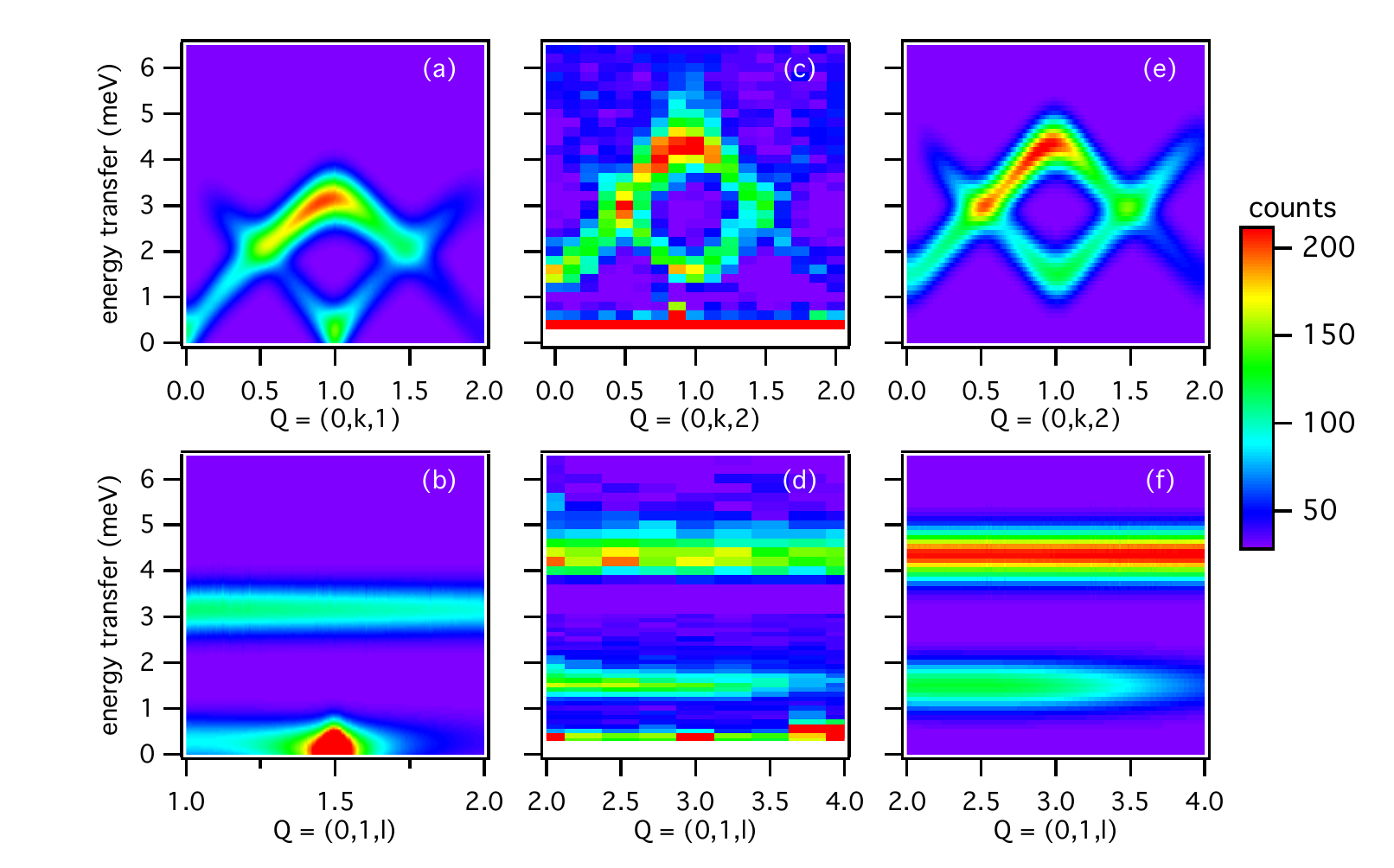}
\caption{Spin wave dispersions for CBSCl in the $(b^*, c^*)$ scattering plane. Simulated dispersion calculated for $\mathbf{Q} = (0,k,1)$ (a) and $\mathbf{Q} = (0,1,l)$ (b) using the proposed Hamiltonian of Rousochatzakis \textit{et al.} \cite{Rousochatzakis2015} (equation~\ref{eq:Hamiltonian}). Measured spin wave dispersion at 2 K for $\mathbf{Q} = (0,k,2)$ (c) and $\mathbf{Q} = (0,1,l)$, mapped from several ThALES energy scans (d). Simulated spin wave dispersion using modified Hamiltonian incorporating symmetric anisotropic exchange for $\mathbf{Q} = (0,k,2)$ (e) and $\mathbf{Q} = (0,1,l)$ (f). Note that the (c-f) maps are indexed in the distorted structure implying a doubling of the $c$ lattice parameter. This results in a doubling of the $l$ index compared to the (a-b) spectra calculated in the undistorted structure.}
\label{fig:THALES_maps}
\end{figure*}

We used a real-space mean-field energy minimization of the spin configuration implemented in the SpinWave software package \cite{SpinWave} to calculate the magnetic ground state in the $Pcmn$ space group. We note that our simulations produce identical results in the $Pmmn$ framework when adjusting for the different $l$ indices. We therefore began with the above Hamiltonian, using the values $J_1$ = -6.55 meV, $J_1'$ = -5.75 meV, $J_2$ = 4.74 meV, $J_{\perp1}$ = -0.035 meV, $J_{\perp2}$ = 0.17 meV and $D_a$ = 1.04 meV stated in reference \cite{Rousochatzakis2015}. Our mean field calculations produce a magnetic structure consistent with the refined $\Gamma_8$ structure depicted in Fig.~\ref{fig:Mag_Struct}.
However, using this form of the Hamiltonian we 
find a smaller Cu1 canting within the $bc$ plane closer to 50\degree as observed in CBSBr. The Cu2 spins are also not stabilized completely along $c$ with some minor canting along $b$ inconsistent with the representations in Table~\ref{tab:MagIrreps}. We note that in reference \cite{Rousochatzakis2015}, values for the $D_b$ and $D_c$ components of the DM interaction have been calculated at $D_b\approx D_c \approx D_a/3$. However, in our simulations we found that while the $D_b$ and $D_c$ components were required to stabilize the magnetic structure, their values could be much smaller without affecting the minimization. In fact at the values stated in reference \cite{Rousochatzakis2015}, the $D_b$ and $D_c$ components had the effect of distorting the calculated spin wave dispersions. We found that a more appropriate relation is given by $D_b\approx D_c \approx D_a/10$. Finally, starting from the ground state and minimizing the energy, we are able to simulate the spin wave spectrum shown in Fig.~\ref{fig:THALES_maps} between 0--6.5 meV for $\mathbf{Q} = (0, k, 1)$ (a) and $\mathbf{Q} = (0, 1, l)$ (b) in good agreement with the simulations of reference \cite{Rousochatzakis2015}.


\begin{figure}
\centering
\includegraphics{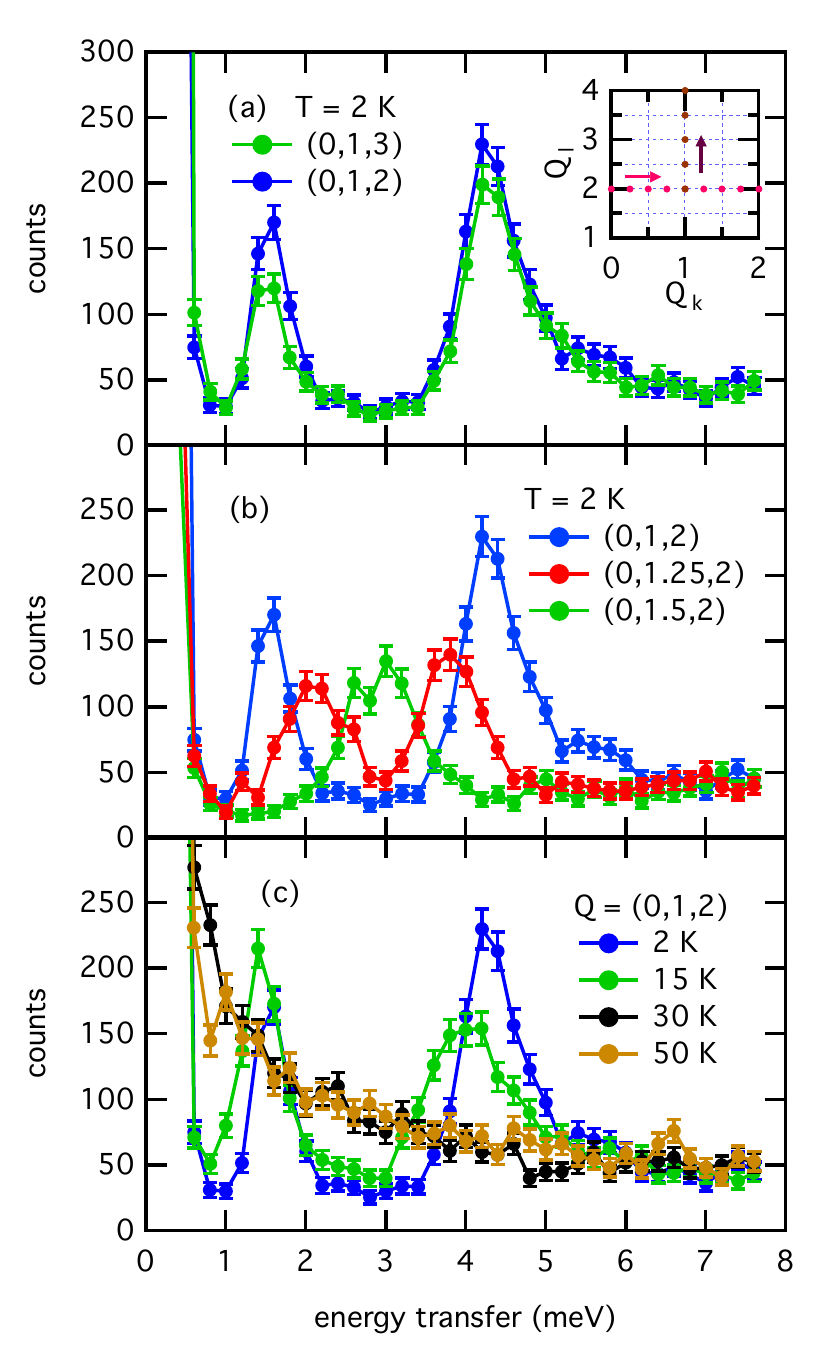}
\caption{Inelastic neutron scattering in the ($b^*$, $c^*$) plane using the ThALES triple axes spectrometer indexed in the low temperature distorted structure. (a) Two energy scans were recorded with energy steps of 0.2 meV at the zone center of different Brillouin zones depicting no significant dispersion along $l$. (b) Selected energy scans for different values of $k$ with $l$ = 2. (c) Temperature evolution of lowest energy excitations at the zone center $(0,1,l)$. The inset in (a) shows the total mapped area in reciprocal space. Data below 0.5 meV corresponds to incoherent elastic scattering.}
\label{fig:THALES_spec}
\end{figure}

We have compared this model to our inelastic neutron scattering measurements. Two series of spin wave spectra were measured in the ($b^*$, $c^*$) scattering plane at 2 K, from 0--8 meV with energy steps of 0.2 meV, along $\mathbf{Q} = (0,k,2)$ and $\mathbf{Q} = (0,1,l)$ with $\Delta Q_h=0.125$  and $\Delta Q_l=0.5$ intervals. Selected spectra are shown in  Fig.~\ref{fig:THALES_spec} (a-d). They are combined into the dispersion maps displayed in Figs.~\ref{fig:THALES_maps} (c) and (d). The spin wave branches below 8 meV are dispersive along $k$ and gapped. At the zone centers, the resulting excitation spectra feature peaks at 1.57 meV and 4.34 meV (see Fig.~\ref{fig:THALES_spec}(a)), consistent with the results of THz \cite{Miller2012} and Raman \cite{Gnezdilov2016_arXiv} spectroscopy. This disagrees however with the spin wave analysis of Rousochatzakis \textit{et al.} \cite{Rousochatzakis2015}, who found gapless excitations at the zone centers, indicating some inaccuracies in the Hamiltonian of equation \ref{eq:Hamiltonian}. The spin waves are dispersionless along $c$ within the energy resolution given by the experimental conditions (see Figs.~\ref{fig:THALES_spec}(a) and ~\ref{fig:THALES_maps}(d)). This highlights the 2-dimensional nature of the kagome planes due to the weakness of the interlayer coupling. When heating to 15 K, both low energy modes at the zone centers shift toward lower frequencies, consistent with temperature dependent Raman scattering spectra \cite{Gnezdilov2016_arXiv}.  
Both modes are completely absent at 30 K and higher temperatures, hence above \TNeel = 25 K, confirming their magnetic origin.

Many of the  key features of these measured spin waves, including the dispersion and the spectral weight of the excitations, 
are captured in the simulations produced by the Hamiltonian of Rousochatzakis \textit{et al.} \cite{Rousochatzakis2015} (equation \ref{eq:Hamiltonian}). However, as mentioned previously, the spin gap is absent. To reproduce this gap we have attempted many modifications to the Hamiltonian including adjusting the interaction strengths, altering the DM interaction vector and introducing anisotropy. We also considered the consequences on the magnetic Hamiltonian of the structural distortion occurring at 115 K from the $Pmmn$ to $Pcmn$ space groups with a doubling of the unit cell along $c$. From our neutron scattering refinements, we find that the Cu1-Cu2 bond distance shifts from a symmetric 3.25 \AA\ at 150 K to either 3.19 \AA\ or 3.32 \AA\ at 2 K. This leads to a splitting of the $J_1'$ interactions into two interactions, $J_{11}'$ for the shorter bonds and $J_{12}'$ for the longer bonds. This is also expected to split the DM interactions. When assessing the refined Hamiltonian, comparisons were made between the mean-field stabilized magnetic structure and our refined structure of Fig~\ref{fig:Mag_Struct}. The simulated spin waves were then compared with the measurements, checking peak positions and intensity over the energy-$Q$ space probed in our experiments. Finally, consideration was also given to whether the adjustments were physically acceptable or not. However, we found little change from these modifications to the simulated magnetic properties, including no induced gap in the spin wave excitations and limited increased canting of Cu1 spins.

More successful attempts to reproduce the observed magnetic structure and the associated spin wave dispersion were actually achieved by introducing an adapted anisotropic exchange. 
The symmetric portion of the anisotropic exchange, that adds to the isotropic exchange interactions of equation \ref{eq:Hamiltonian}, is implemented into the Hamiltonian with the following term,
\begin{equation}
H_\mathrm{AE} =  \mathbf{S}_i\cdot \mathbf{\Gamma}_{i,j} \cdot \mathbf{S}_j,
\end{equation}
where $\mathbf{\Gamma}_{i,j}$ is a second-rank symmetric tensor.
The tensors $\mathbf{\Gamma}_{1,2}$, $\mathbf{\Gamma}_{5,2}$ and $\mathbf{\Gamma}_{3,2}$ then represent the symmetric anisotropic exchange for the bonds between the sites labeled in Fig.\ref{fig:Mag_Struct} (c) (as well as all other symmetrically equivalent bonds). In their calculations, Rousochatzakis \textit{et al.} found that the contribution of this symmetric anisotropic exchange was negligible \cite{Rousochatzakis2015}. However, in our analysis we have found  that the combination of small positive (antiferromagnetic) off-diagonal  $\Gamma^{yz}_{1,2}$ and $\Gamma^{zy}_{1,2}$ components and larger negative components for $\Gamma^{xx}_{4,2}$ and  $\Gamma^{zz}_{4,2}$ is able to open the gap in the spin wave spectrum. Further adjustment of the gap energy and of the global dispersion, as well as the canting of the Cu1 spins, is achieved by including a small component for $\Gamma^{xx}_{5,2}$ and by slightly changing the other terms of the Hamiltonian. %
Finally, we find an adequate solution by only a slight increase in the DM interaction accompanied by an adjustment of $J_2/\left|J_1'\right| \approx 1 $, yielding the following modified magnetic Hamiltonian with
$J_1$ = -6.55 meV, $J_1'$ = -5.75 meV, $J_2$ = 5.7 meV, $J_{\perp1}$ = -0.035 meV, $J_{\perp2}$ = 0.17 meV and $D_a$ = 1.3 meV, with $D_b=D_c=D_a/10$, and the following symmetric portion of the anisotropic exchange with values represented in meV:
\begin{eqnarray}
\Gamma_{1,2} &=& \left(\begin{matrix}
									0.00&0.00&0.00\\
									0.00&0.05&\underline{0.12}\\
									0.00&\underline{0.12}&0.05\\
\end{matrix}\right), \nonumber\\*
\Gamma_{5,2} &=& \left(\begin{matrix}
									\underline{0.10}&0.08&0.00\\
									0.08&0.00&0.04\\
									0.00&0.04&0.00\\
\end{matrix}\right), \nonumber\\*
\Gamma_{4,2} &=& \left(\begin{matrix}
									\underline{-0.7}&0.0&0.0\\
									0.0&0.0&0.0\\
									0.0&0.0&\underline{-1.7}\\
\end{matrix}\right). \nonumber
\end{eqnarray}
The terms that are significant for generating the spin wave gap are underlined. Note that the anisotropic exchange tensors are essentially equivalent to those calculated in Ref. \cite{Rousochatzakis2015} with the exception of the relatively large components we have found for $\mathbf{\Gamma}_{4,2}$. We note that there is a high degree of uncertainty associated to the refined values in $\mathbf{\Gamma}_{4,2}$. Taking the presence of a spin wave gap and a $\sim 3$ meV splitting of the two spin wave branches at $k=1$ as refinement limits, we find an approximate upper boundary for the uncertainties of these components to be $\mathbf{\Gamma}^{xx}_{4,2}=-0.7\pm0.4$ and $\mathbf{\Gamma}^{zz}_{4,2}=-1.7\pm1.0$.

By incorporating this anisotropic exchange into the Hamiltonian, we are thus able to reproduce the magnetic structure with no instability along $b$ for the Cu2 spins. The calculated canting of the Cu1 spins is 59\degree from $c$ to $a$ in close agreement with our single crystal neutron diffraction refinement. No canting is found along the $a$ direction. The simulated spin wave dispersions using the modified Hamiltonian  for $\mathbf{Q}=(0, k, 2)$ and $\mathbf{Q}=(0, 1, l)$ are shown in Figs.~\ref{fig:THALES_maps} (e) and (f) respectively. Here we find excellent agreement between experiment (Figs.~\ref{fig:THALES_maps} (c), (d)) and simulation (Figs.~\ref{fig:THALES_maps} (e), (f))  when comparing the intensity and dispersions along $k$ and $l$. Our INS experiments thus reveal that a next-nearest-neighbor exchange between the Cu1 sites in the $ac$ plane includes a non-negligible anisotropic exchange part. This then induces the gapped spin wave spectrum we observe.

\begin{figure}
\centering
\includegraphics{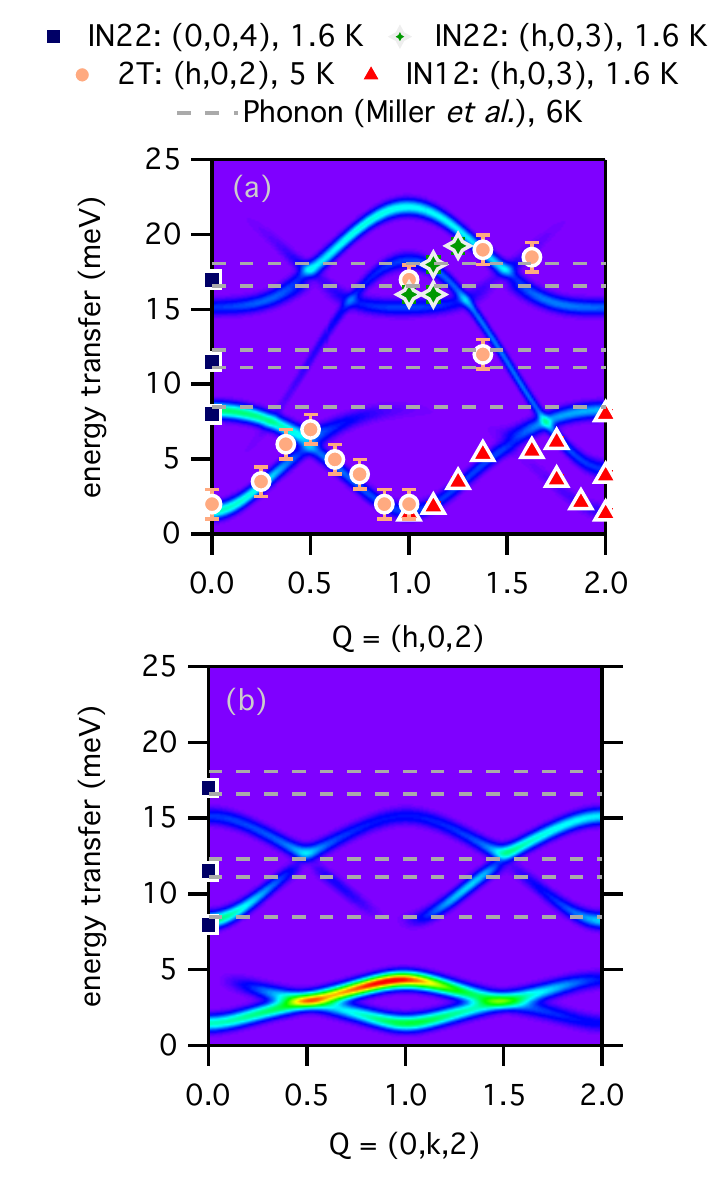}
\caption{ 
(a) Simulated spin wave dispersion for $\mathbf{Q} = (h, 0, 2)$ using the modified Hamiltonian. The peak positions of the excitations measured using the IN12, IN22 and 2T-1 triple-axis spectrometers are plotted on top of the simulation. Gray dashed lines indicate the locations of transverse optical phonons measured by infrared spectroscopy \cite{Miller2012}. 
(b) Simulated spin wave dispersion for $\mathbf{Q} = (0,k,2)$ using the modified Hamiltonian. Extracted peak positions (symbols) and phonon energies (gray dashed lines) are plotted on top.}
\label{fig:in22phns}
\end{figure}

Further agreement between experiment and simulation is obtained for the ($a^*$, $c^*$) scattering plane.
As is predicted in our simulations, the scattering intensity of the spin waves for the ($a^*$, $c^*$) scattering plane are reduced by a factor of $\sim$3 when compared to the ($b^*$, $c^*$) scattering plane, rendering it more difficult to map the dispersion in this orientation. Nevertheless, by combining measurements performed on several spectrometers, we were able to extract peak positions across a broad span of $h$ between 0 and 20 meV. The resulting experimental and simulated dispersion is shown in Fig.~\ref{fig:in22phns} (a). The extracted peak positions, plotted on top of the simulation using the modified Hamiltonian, show a good agreement at energies below 8 meV. Fig.~\ref{fig:in22phns} (b) reveals the predicted dispersion of higher energy spin waves in the ($b^*$, $c^*$) scattering plane including extracted peak positions above 8 meV. The energies of transverse optical phonons observed by infrared spectroscopy \cite{Miller2012} are indicated by dashed gray lines plotted over the simulations.

Concerning the higher energy spin wave branches in both the ($a^*$, $c^*$) and ($b^*$, $c^*$) scattering planes, no significant magnetic contribution could be detected in the INS experiments when comparing spectra above and below \TNeel. 
In fact, while many of the higher energy peaks (above 8 meV) measured in our INS experiments are aligned with predicted spin wave branches, they are also in close proximity to the previously observed infrared optical phonons. The presence of spin-wave--phonon hybridization has been previously established in CBSCl \cite{Gnezdilov2016_arXiv}. It is therefore possible that the higher energy spin wave branches are either masked or hybridized with phonon signatures observed in Fig.\ref{fig:in22phns}. Further experiments incorporating polarized neutron analysis will be required to investigate this possibility.

\subsection{Dielectric and polarization properties}

\begin{figure}
\centering
\includegraphics{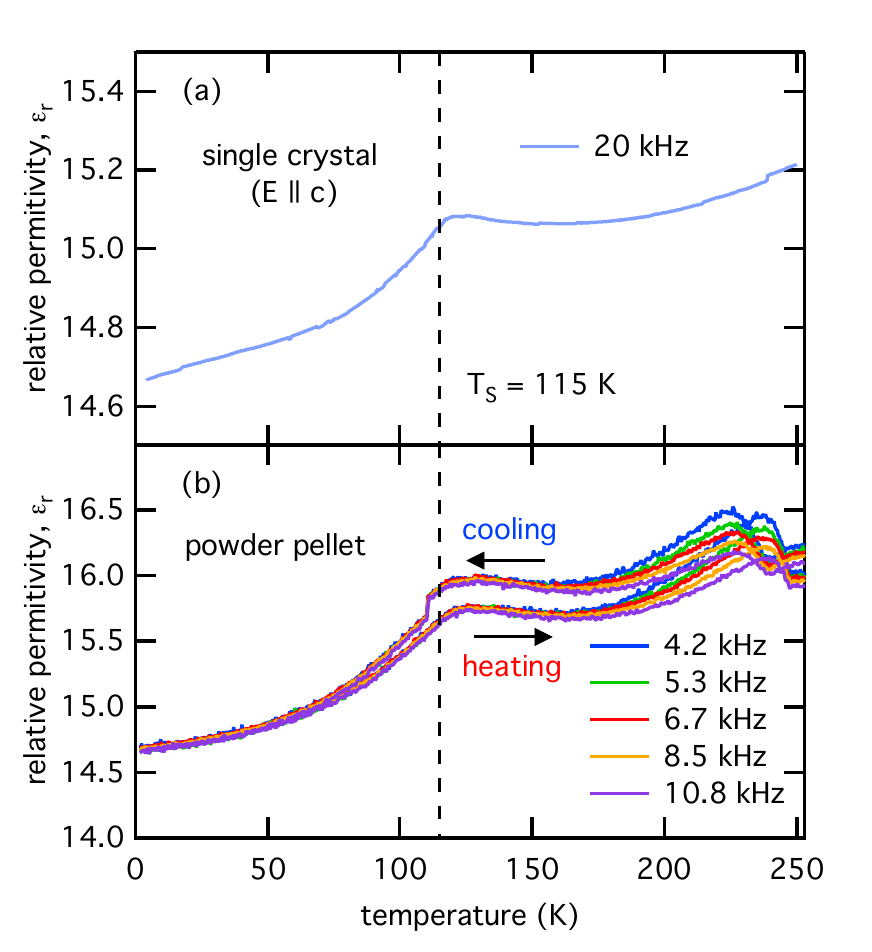}
\caption{ (a) Relative permittivity of single crystal along the $c$ axis (a) and powder pellet (b) CBSCl samples as a function of temperature and for various frequencies in the range 2--10 kHz. A bias of 1 V was used in the measurements.}
\label{fig:Perm}
\end{figure}

To further study the phase diagram of CBSCl, we have looked at possible signatures of the structural and magnetic transitions in the dielectric properties.
Electrical capacitative measurements were used to determine the relative permittivity of CBSCl shown in Figs. \ref{fig:Perm} (a) and (b) for a single crystal along the $c$ axis and for a powder pellet sample respectively. In Fig. \ref{fig:Perm} (a), a dielectric anomaly is observed at $\sim$120 K, coinciding with the structural transition. The anomaly is continuous, which is characteristic of a 2$^{\rm nd}$-order phase transition. At the dielectric anomaly, we observe no frequency dependence, thus indicating long-range polar correlations, nor thermal hysteresis suggesting no kinetic effects. Since we have shown that the structural transition involves a distortion from the centrosymmetric $Pmmn$ space group to the centrosymetric $Pcmn$ space group, we rule out the formation of long-range polar order into a ferroelectric (FE) phase and turn to the possibility of an antiferroelectric (AFE) low temperature phase.

Antiferroelectrics are a somewhat loosely defined class of non-polar materials. They are typically distinguished from standard non-polar systems at the microscopic scale by considering the presence of non-collinear dipoles mutually canceling \cite{Toledano_2016}. They tend to form following a distortion from one symmetric phase to a less symmetric phase accompanied by an anomaly in the dielectric permittivity related to the formation and/or organization of local dipole moments in an AF arrangement. This can also involve a doubling of the cell parameters when the associated soft phonon modes feature two groups of displacements, equal and opposite in magnitude \cite{Rabe_2013,Devonshire_1954}. The basic theoretical description of these AFEs is therefore one with two (or multiples of two) equivalent and opposing polarized sublattices \cite{Kittel_1951}. The application of a sufficiently large electric field along the dipolar axis is expected to invert the polarization of the opposing sublattice, forming an induced ferroelectric state, and leading to a characteristic hysteresis loop in the electrical polarization as a function of the electric field \cite{Toledano_2016,Rabe_2013}. The $Pcmn$ phase in CBSCl is consistent with the transition at \TStr from a paralelectric (PE) phase to an AFE state as described above. The primary distortion involves displacements of the Cu2 and Cl ions in opposing directions along $a$ giving rise to a doubling of the $c$ cell parameter. Note that the isotropic displacement parameter of the Cl$^-$ ion obtained from the neutron refinement (see Table \ref{tab:Atms}) is relatively large, which can indicate a high degree of mobility. This distortion is accompanied by the anomaly in the dielectric permittivity and provides clear candidates for opposing dipoles lying primarily within the $ab$ plane.

The thin planar geometry of the single crystal samples confines electrical measurements along the $c$ axis. To probe the dielectric response in the $ab$ plane, we look at measurements on the powder pellet sample, averaged over all crystallographic directions. As evident in Fig. \ref{fig:Perm} (b), the same dielectric anomaly at 120 K is observed in the powder pellet sample signifying the formation of the AFE phase. Again, no frequency dependence is present. However,  in contrast to the single crystal results, a second anomaly is observed at higher temperatures around 230 K displaying thermal hysteresis as well as a small frequency dependence.
The origin of this feature could arise from different dynamics for highly mobile electric charges present in the material, for instance the Cl$^-$ ions, which are impeded by the many grain boundaries of the pellet sample but not for the single crystal. 

\begin{figure}
\centering
\includegraphics{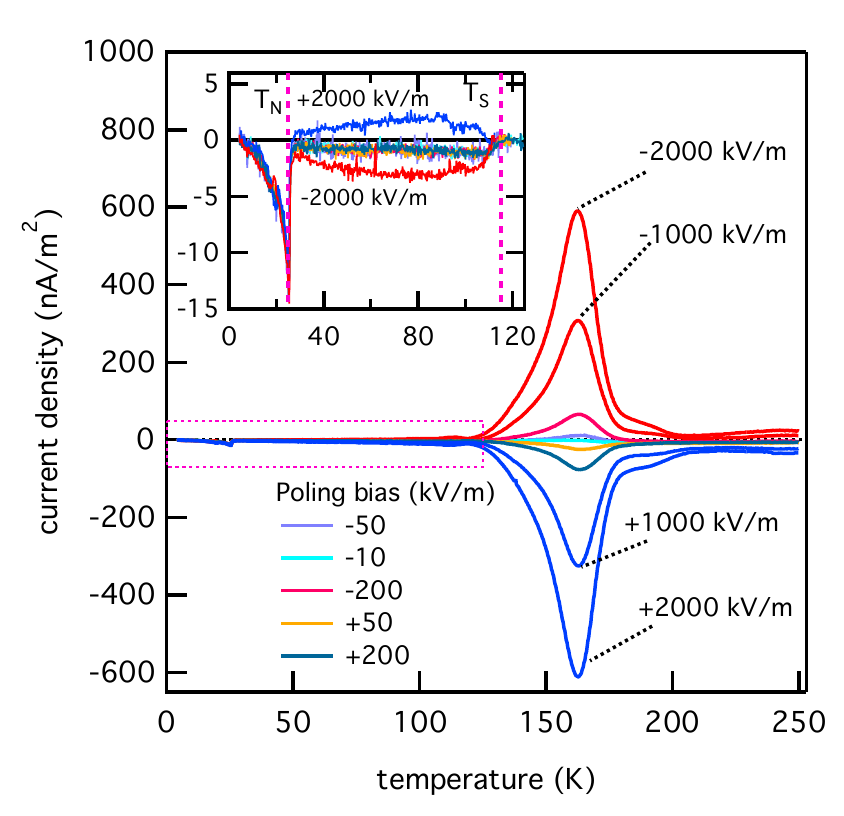}
\caption{Time dependent depolarizing current for single crystal CBSCl along $c$ measured at a heating rate of 3 K/min. The results are shown as current density over temperature and reveal a large depolarizing current released above \TStr. The inset shows a zoomed-in picture of the current measured between 4 and 125 K with small features at \TNeel and \TStr. A linear background dependence has be removed from the data presented in the inset.}
\label{fig:Pyro}
\end{figure}

To provide further information regarding the high temperature dispersive anomaly of Fig.~\ref{fig:Perm} (b) and the nature of the PE-AFE transition in general, we performed additional electric polarization measurements. The temperature dependent depolarizing current measurements used to calculate the electrical polarization are shown in Fig. \ref{fig:Pyro} for the single crystal sample.
A large current is detected above \TStr forming a peak around 160 K. It follows a linear dependence on the poling bias and is flipped when the polarization of the bias is reversed. It is related to the dispersive dielectric anomaly discussed above, which appears at higher temperatures but shifts to lower temperatures at lower frequencies, consistent with the occurrence at 160 K of the current peak, which can be considered a DC measurement. The released current is detected in both the single crystal and powder pellet sample (not shown), which suggests that the process is intrinsic to the material and does not depend on the sample form. This depolarizing current may be the result of trapped charges that are freed when microscopic AFE domains relax in the unstable thermal region above \TStr. Their contributions to the pyroelectric current are removed when considering calculations of the electrical polarization.

\begin{figure}
\centering
\includegraphics{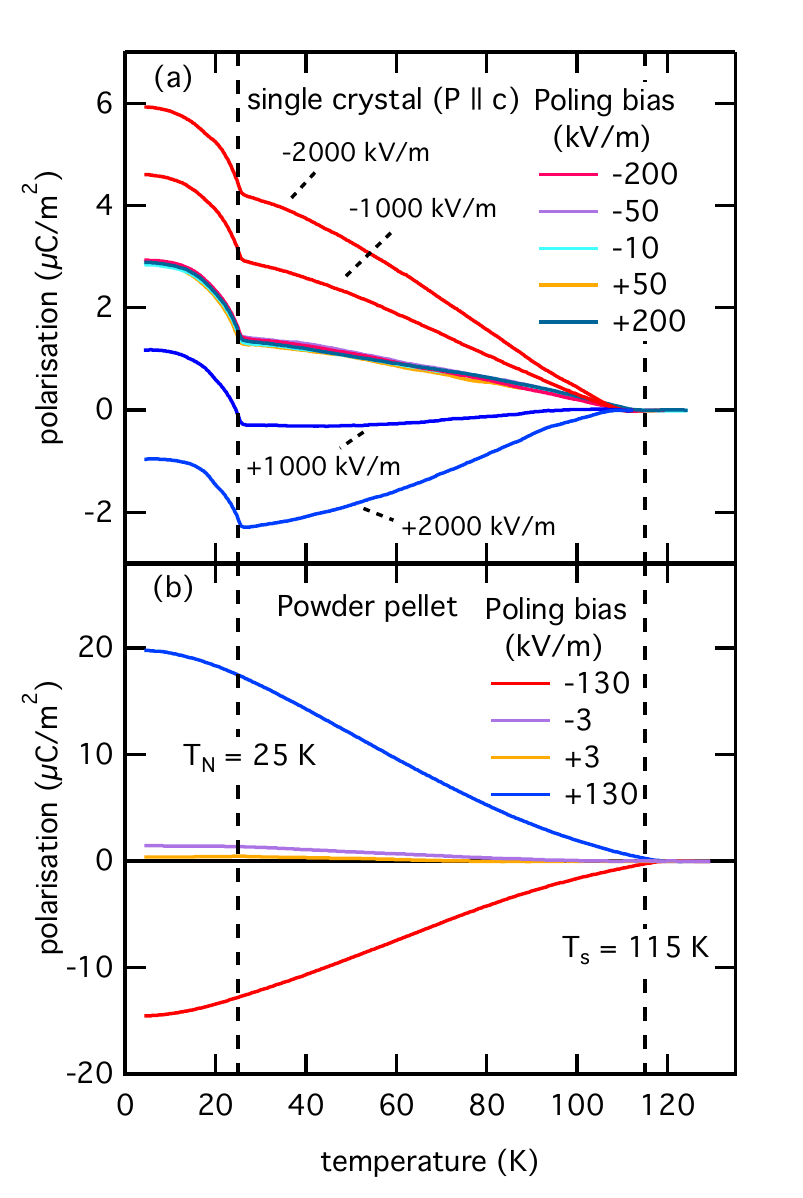}
\caption{Calculated polarization as a function of temperature between 125 and 4 K, from the integration of the pyroelectric current, under multiple electric poling fields for a single crystal of CBSCl with the electric field applied along the $c$ axis (a) and for a powder pellet sample (b).}
\label{fig:Pol}
\end{figure}

The pyroelectric current, measured for an electric field applied along the $c$ axis on a single-crystal, is shown in the inset of figure \ref{fig:Pyro} for the low temperature phases. Two small features are detected, one at the structural transition (\TStr) and one at the magnetic transition (\TNeel). The resulting electrical polarization over this region is shown in Fig. \ref{fig:Pol}. A small electrical polarization is detected below \TStr, increasing at lower temperature, with an additional small contribution occurring below \TNeel. At electric fields up to $\pm$200 kV/m, neither the contribution below \TStr nor the contribution below \TNeel are affected by the poling bias. At higher electric fields in the range of $\pm$1000 kV/m, the polarization is manipulated below \TStr, switching polarity between the application of positive and negative poling fields. However, the contribution to the polarization below \TNeel remains unaffected, even at the highest poling fields achievable in our experimental setup. The polarization obtained for the powder pellet sample is shown in Fig. \ref{fig:Pol} (b). The polarization increases for decreasing temperature, and is significantly larger than that detected for the single crystal along the $c$ axis under equivalent poling fields. Under small poling fields a negligible polarization is observed, likely only displaying a contribution from the residual polarization detected along the $c$ axis. Under larger poling fields, the polarization is increased and is switchable with the polarity of the field. Notably, the polarization contribution below \TNeel detected along the $c$ axis is not observed in the powder pellet measurement.

In order to interpret these measurements, we recall that the structural distortion at 115 K involves a transformation between two centrosymmetric space groups, $Pmmn$ and $Pcmn$, that do not support macroscopic polar moments. Actually, the polarization switching in the polycrystalline sample and at high poling fields in the single-crystal is consistent with an induced polarization in an AFE phase, and with the notion of antiparallel dipoles in the $ab$ plane due to the displacements of the Cu2 and Cl ions along $a$. 

Another perplexing result is the presence of the weak non-switchable polarization at \TNeel with no signature in the dielectric permittivity. It is characteristic of some form of magnetoelectric coupling, and only detected in the single-crystal. The fact that is is not detected in the powder sample suggests that the magnetoelectric polarization is indeed along the $c$ axis. 
The magnetic point group $mm'm$ that describes the magnetic phase of CBSCl is compatible with a linear magnetoelectric coupling, $\alpha^{ME}_{ij}E_iH_j$, with $E_i$ and $H_j$ the components of the electric and magnetic fields respectively and $\alpha^{ME}_{ij}$ the components of the magnetoelectric tensor of the form \cite{IntTbs_Crst_D_2013}: 
\begin{equation}
\alpha^{ME} = \left(\begin{matrix}
0&0&\alpha^{ME}_{13}\\
0&0&0\\
\alpha^{ME}_{31}&0&0
\end{matrix}\right).
\end{equation}
The off-diagonal terms indicate that the coupling is maximized when the electric and magnetic fields are perpendicular. Its form also allows ferrotoroidicity as the toroidic moment is proportional to the antisymmetric part of the magnetoelectric tensor \cite{Gorbatsevich_1983}. 
However in the absence of a uniform magnetic field, it should not lead to a net macroscopic polarization. This anomalous electric polarization we observe may rather be associated to symmetry breaking of local magnetic fields generated at defect or domain wall boundaries.
A recent study of the multiferroic behavior in CBSCl has indeed shown that an anisotropic spin-flip transition induced by magnetic fields above $\sim$0.8 T leads to a breaking of the magnetic two-fold symmetry in the $bc$ plane \cite{Wu2017}. The resulting phase demonstrates ferroelectric behavior below \TNeel.
\section{Conclusion \label{sec:Concltn}}

By performing single crystal neutron diffraction measurements on the CBSCl francisite, we confirm the structural phase transition at 115 K as a distortion of the high temperature $Pmmn$ phase to a $Pcmn$ phase with a doubling of the $c$ cell parameter. Within the distorted low temperature structure, we refine the true magnetic structure of CBSCl for the first time revealing a structure similar to that of CBSBr but with an increased canting of 59\degree for Cu1 spins. Inelastic neutron scattering experiments have shown a global energy gap in spin wave spectrum, which is inconsistent with the previously established form of the Hamiltonian describing CBSCl. Refinement of the Hamiltonian through mean-field minimization and spin wave simulations reveals that a non-negligible symmetric exchange anisotropy is required, on top of the weak Dzyaloshinski-Moriya interaction, to stabilize the magnetic phase and induce the spin wave gap.
From electric measurements (dielectric response and electric polarization), we show that the structural transition induces an antiferroelectric phase below 115 K and point out the existence of a transverse magnetoelectric coupling below \TNeel. However, the precise 
origin of the measured weak polarization in zero magnetic field 
still remains to be determined.
Together the results deepen the understanding of this complex material and highlight the interesting functional capabilities of the whole material class.


\section {Acknowledgements}

 We would like to thank L. Del-Rey for his help in the design of the neutron sample-holder, E. Lefrancois for her help on CYCLOPS, M. Boehm and P. Steffens for their support on ThALES, M. Aroyo for his advice on the symmetry analysis, R. Ballou for fruitful discussions, and A. Cano for coordinating the dielectric experiments carried out in Bordeaux. We also acknowledge K. H. Miller and D. B. Tanner for the sample acquisition, and F. Gay for his contribution to the polarization experimental setup. M. Josse would like to thank Dr. Mario Maglione for useful discussion. This work was financially supported by Grant No. ANR-13-BS04-0013. E. Constable has benefited from a PRESTIGE (No.2014-1-0020) fellowship and CMIRA funding for this research work.


\begin{thebibliography}{31}%
\makeatletter
\providecommand \@ifxundefined [1]{%
 \@ifx{#1\undefined}
}%
\providecommand \@ifnum [1]{%
 \ifnum #1\expandafter \@firstoftwo
 \else \expandafter \@secondoftwo
 \fi
}%
\providecommand \@ifx [1]{%
 \ifx #1\expandafter \@firstoftwo
 \else \expandafter \@secondoftwo
 \fi
}%
\providecommand \natexlab [1]{#1}%
\providecommand \enquote  [1]{``#1''}%
\providecommand \bibnamefont  [1]{#1}%
\providecommand \bibfnamefont [1]{#1}%
\providecommand \citenamefont [1]{#1}%
\providecommand \href@noop [0]{\@secondoftwo}%
\providecommand \href [0]{\begingroup \@sanitize@url \@href}%
\providecommand \@href[1]{\@@startlink{#1}\@@href}%
\providecommand \@@href[1]{\endgroup#1\@@endlink}%
\providecommand \@sanitize@url [0]{\catcode `\\12\catcode `\$12\catcode
  `\&12\catcode `\#12\catcode `\^12\catcode `\_12\catcode `\%12\relax}%
\providecommand \@@startlink[1]{}%
\providecommand \@@endlink[0]{}%
\providecommand \url  [0]{\begingroup\@sanitize@url \@url }%
\providecommand \@url [1]{\endgroup\@href {#1}{\urlprefix }}%
\providecommand \urlprefix  [0]{URL }%
\providecommand \Eprint [0]{\href }%
\providecommand \doibase [0]{http://dx.doi.org/}%
\providecommand \selectlanguage [0]{\@gobble}%
\providecommand \bibinfo  [0]{\@secondoftwo}%
\providecommand \bibfield  [0]{\@secondoftwo}%
\providecommand \translation [1]{[#1]}%
\providecommand \BibitemOpen [0]{}%
\providecommand \bibitemStop [0]{}%
\providecommand \bibitemNoStop [0]{.\EOS\space}%
\providecommand \EOS [0]{\spacefactor3000\relax}%
\providecommand \BibitemShut  [1]{\csname bibitem#1\endcsname}%
\let\auto@bib@innerbib\@empty
\bibitem [{\citenamefont {Chubukov}(1992)}]{Chubukov_1992}%
  \BibitemOpen
  \bibfield  {author} {\bibinfo {author} {\bibfnamefont {A.}~\bibnamefont
  {Chubukov}},\ }\href {\doibase 10.1103/physrevlett.69.832} {\bibfield
  {journal} {\bibinfo  {journal} {Physical Review Letters}\ }\textbf {\bibinfo
  {volume} {69}},\ \bibinfo {pages} {832} (\bibinfo {year} {1992})}\BibitemShut
  {NoStop}%
\bibitem [{\citenamefont {Sachdev}(1992)}]{Sachdev_1992}%
  \BibitemOpen
  \bibfield  {author} {\bibinfo {author} {\bibfnamefont {S.}~\bibnamefont
  {Sachdev}},\ }\href {\doibase 10.1103/physrevb.45.12377} {\bibfield
  {journal} {\bibinfo  {journal} {Physical Review B}\ }\textbf {\bibinfo
  {volume} {45}},\ \bibinfo {pages} {12377} (\bibinfo {year}
  {1992})}\BibitemShut {NoStop}%
\bibitem [{\citenamefont {Shores}\ \emph {et~al.}(2005)\citenamefont {Shores},
  \citenamefont {Nytko}, \citenamefont {Bartlett},\ and\ \citenamefont
  {Nocera}}]{Shores_2005}%
  \BibitemOpen
  \bibfield  {author} {\bibinfo {author} {\bibfnamefont {M.~P.}\ \bibnamefont
  {Shores}}, \bibinfo {author} {\bibfnamefont {E.~A.}\ \bibnamefont {Nytko}},
  \bibinfo {author} {\bibfnamefont {B.~M.}\ \bibnamefont {Bartlett}}, \ and\
  \bibinfo {author} {\bibfnamefont {D.~G.}\ \bibnamefont {Nocera}},\ }\href
  {\doibase 10.1021/ja053891p} {\bibfield  {journal} {\bibinfo  {journal}
  {Journal of the American Chemical Society}\ }\textbf {\bibinfo {volume}
  {127}},\ \bibinfo {pages} {13462} (\bibinfo {year} {2005})}\BibitemShut
  {NoStop}%
\bibitem [{\citenamefont {Helton}\ \emph {et~al.}(2007)\citenamefont {Helton},
  \citenamefont {Matan}, \citenamefont {Shores}, \citenamefont {Nytko},
  \citenamefont {Bartlett}, \citenamefont {Yoshida}, \citenamefont {Takano},
  \citenamefont {Suslov}, \citenamefont {Qiu}, \citenamefont {Chung},
  \citenamefont {Nocera},\ and\ \citenamefont {Lee}}]{Helton_2007}%
  \BibitemOpen
  \bibfield  {author} {\bibinfo {author} {\bibfnamefont {J.~S.}\ \bibnamefont
  {Helton}}, \bibinfo {author} {\bibfnamefont {K.}~\bibnamefont {Matan}},
  \bibinfo {author} {\bibfnamefont {M.~P.}\ \bibnamefont {Shores}}, \bibinfo
  {author} {\bibfnamefont {E.~A.}\ \bibnamefont {Nytko}}, \bibinfo {author}
  {\bibfnamefont {B.~M.}\ \bibnamefont {Bartlett}}, \bibinfo {author}
  {\bibfnamefont {Y.}~\bibnamefont {Yoshida}}, \bibinfo {author} {\bibfnamefont
  {Y.}~\bibnamefont {Takano}}, \bibinfo {author} {\bibfnamefont
  {A.}~\bibnamefont {Suslov}}, \bibinfo {author} {\bibfnamefont
  {Y.}~\bibnamefont {Qiu}}, \bibinfo {author} {\bibfnamefont {J.-H.}\
  \bibnamefont {Chung}}, \bibinfo {author} {\bibfnamefont {G.~G.}\ \bibnamefont
  {Nocera}}, \ and\ \bibinfo {author} {\bibfnamefont {Y.~S.}\ \bibnamefont
  {Lee}},\ }\href {http://dx.doi.org/10.1103/PhysRevLett.98.107204} {\bibfield
  {journal} {\bibinfo  {journal} {Physical Review Letters}\ }\textbf {\bibinfo
  {volume} {98}},\ \bibinfo {pages} {107204} (\bibinfo {year}
  {2007})}\BibitemShut {NoStop}%
\bibitem [{\citenamefont {Yan}\ \emph {et~al.}(2011)\citenamefont {Yan},
  \citenamefont {Huse},\ and\ \citenamefont {White}}]{Yan_2011}%
  \BibitemOpen
  \bibfield  {author} {\bibinfo {author} {\bibfnamefont {S.}~\bibnamefont
  {Yan}}, \bibinfo {author} {\bibfnamefont {D.~A.}\ \bibnamefont {Huse}}, \
  and\ \bibinfo {author} {\bibfnamefont {S.~R.}\ \bibnamefont {White}},\ }\href
  {\doibase 10.1126/science.1201080} {\bibfield  {journal} {\bibinfo  {journal}
  {Science}\ }\textbf {\bibinfo {volume} {332}},\ \bibinfo {pages} {1173}
  (\bibinfo {year} {2011})}\BibitemShut {NoStop}%
\bibitem [{\citenamefont {Balents}(2010)}]{Balents_2010}%
  \BibitemOpen
  \bibfield  {author} {\bibinfo {author} {\bibfnamefont {L.}~\bibnamefont
  {Balents}},\ }\href {http://dx.doi.org/10.1038/nature08917} {\bibfield
  {journal} {\bibinfo  {journal} {Nature}\ }\textbf {\bibinfo {volume} {464}},\
  \bibinfo {pages} {199} (\bibinfo {year} {2010})}\BibitemShut {NoStop}%
\bibitem [{\citenamefont {Bieri}\ \emph {et~al.}(2015)\citenamefont {Bieri},
  \citenamefont {Messio}, \citenamefont {Bernu},\ and\ \citenamefont
  {Lhuillier}}]{Bieri_2015}%
  \BibitemOpen
  \bibfield  {author} {\bibinfo {author} {\bibfnamefont {S.}~\bibnamefont
  {Bieri}}, \bibinfo {author} {\bibfnamefont {L.}~\bibnamefont {Messio}},
  \bibinfo {author} {\bibfnamefont {B.}~\bibnamefont {Bernu}}, \ and\ \bibinfo
  {author} {\bibfnamefont {C.}~\bibnamefont {Lhuillier}},\ }\href
  {http://dx.doi.org/10.1103/PhysRevB.92.060407} {\bibfield  {journal}
  {\bibinfo  {journal} {Physical Review B}\ }\textbf {\bibinfo {volume} {92}},\
  \bibinfo {pages} {060407(R)} (\bibinfo {year} {2015})}\BibitemShut {NoStop}%
\bibitem [{\citenamefont {F{\aa}k}\ \emph {et~al.}(2012)\citenamefont
  {F{\aa}k}, \citenamefont {Kermarrec}, \citenamefont {Messio}, \citenamefont
  {Bernu}, \citenamefont {Lhuillier}, \citenamefont {Bert}, \citenamefont
  {Mendels}, \citenamefont {Koteswararao}, \citenamefont {Bouquet},
  \citenamefont {Ollivier}, \citenamefont {Hiller}, \citenamefont {Amato},
  \citenamefont {Colman},\ and\ \citenamefont {Wills}}]{Fak_2012}%
  \BibitemOpen
  \bibfield  {author} {\bibinfo {author} {\bibfnamefont {B.}~\bibnamefont
  {F{\aa}k}}, \bibinfo {author} {\bibfnamefont {E.}~\bibnamefont {Kermarrec}},
  \bibinfo {author} {\bibfnamefont {L.}~\bibnamefont {Messio}}, \bibinfo
  {author} {\bibfnamefont {B.}~\bibnamefont {Bernu}}, \bibinfo {author}
  {\bibfnamefont {C.}~\bibnamefont {Lhuillier}}, \bibinfo {author}
  {\bibfnamefont {F.}~\bibnamefont {Bert}}, \bibinfo {author} {\bibfnamefont
  {P.}~\bibnamefont {Mendels}}, \bibinfo {author} {\bibfnamefont
  {B.}~\bibnamefont {Koteswararao}}, \bibinfo {author} {\bibfnamefont
  {F.}~\bibnamefont {Bouquet}}, \bibinfo {author} {\bibfnamefont
  {J.}~\bibnamefont {Ollivier}}, \bibinfo {author} {\bibfnamefont {A.~D.}\
  \bibnamefont {Hiller}}, \bibinfo {author} {\bibfnamefont {A.}~\bibnamefont
  {Amato}}, \bibinfo {author} {\bibfnamefont {R.~H.}\ \bibnamefont {Colman}}, \
  and\ \bibinfo {author} {\bibfnamefont {A.~S.}\ \bibnamefont {Wills}},\ }\href
  {http://dx.doi.org/10.1103/PhysRevLett.109.037208} {\bibfield  {journal}
  {\bibinfo  {journal} {Physical Review Letters}\ }\textbf {\bibinfo {volume}
  {109}},\ \bibinfo {pages} {037208} (\bibinfo {year} {2012})}\BibitemShut
  {NoStop}%
\bibitem [{\citenamefont {Nilsen}\ \emph {et~al.}(2014)\citenamefont {Nilsen},
  \citenamefont {Okamoto}, \citenamefont {Ishikawa}, \citenamefont {Simonet},
  \citenamefont {Colin}, \citenamefont {Cano}, \citenamefont {Chapon},
  \citenamefont {Hansen}, \citenamefont {Mutka},\ and\ \citenamefont
  {Hiroi}}]{Nilsen_2014}%
  \BibitemOpen
  \bibfield  {author} {\bibinfo {author} {\bibfnamefont {G.~J.}\ \bibnamefont
  {Nilsen}}, \bibinfo {author} {\bibfnamefont {Y.}~\bibnamefont {Okamoto}},
  \bibinfo {author} {\bibfnamefont {H.}~\bibnamefont {Ishikawa}}, \bibinfo
  {author} {\bibfnamefont {V.}~\bibnamefont {Simonet}}, \bibinfo {author}
  {\bibfnamefont {C.~V.}\ \bibnamefont {Colin}}, \bibinfo {author}
  {\bibfnamefont {A.}~\bibnamefont {Cano}}, \bibinfo {author} {\bibfnamefont
  {L.~C.}\ \bibnamefont {Chapon}}, \bibinfo {author} {\bibfnamefont
  {T.}~\bibnamefont {Hansen}}, \bibinfo {author} {\bibfnamefont
  {H.}~\bibnamefont {Mutka}}, \ and\ \bibinfo {author} {\bibfnamefont
  {Z.}~\bibnamefont {Hiroi}},\ }\href
  {http://dx.doi.org/10.1103/PhysRevB.89.140412} {\bibfield  {journal}
  {\bibinfo  {journal} {Physical Review B}\ }\textbf {\bibinfo {volume} {89}},\
  \bibinfo {pages} {140412(R)} (\bibinfo {year} {2014})}\BibitemShut {NoStop}%
\bibitem [{\citenamefont {Pring}\ \emph {et~al.}(1990)\citenamefont {Pring},
  \citenamefont {Gatehouse},\ and\ \citenamefont {Birch}}]{Pring1990}%
  \BibitemOpen
  \bibfield  {author} {\bibinfo {author} {\bibfnamefont {A.}~\bibnamefont
  {Pring}}, \bibinfo {author} {\bibfnamefont {B.~M.}\ \bibnamefont
  {Gatehouse}}, \ and\ \bibinfo {author} {\bibfnamefont {W.~D.}\ \bibnamefont
  {Birch}},\ }\href@noop {} {\bibfield  {journal} {\bibinfo  {journal}
  {American Mineralogist}\ }\textbf {\bibinfo {volume} {75}},\ \bibinfo {pages}
  {1421} (\bibinfo {year} {1990})}\BibitemShut {NoStop}%
\bibitem [{\citenamefont {Millet}\ \emph {et~al.}(2001)\citenamefont {Millet},
  \citenamefont {Bastide}, \citenamefont {Pashchenko}, \citenamefont
  {Gnatchenko}, \citenamefont {Gapon}, \citenamefont {Ksari},\ and\
  \citenamefont {Stepanov}}]{Millet2001}%
  \BibitemOpen
  \bibfield  {author} {\bibinfo {author} {\bibfnamefont {P.}~\bibnamefont
  {Millet}}, \bibinfo {author} {\bibfnamefont {B.}~\bibnamefont {Bastide}},
  \bibinfo {author} {\bibfnamefont {V.}~\bibnamefont {Pashchenko}}, \bibinfo
  {author} {\bibfnamefont {S.}~\bibnamefont {Gnatchenko}}, \bibinfo {author}
  {\bibfnamefont {V.}~\bibnamefont {Gapon}}, \bibinfo {author} {\bibfnamefont
  {Y.}~\bibnamefont {Ksari}}, \ and\ \bibinfo {author} {\bibfnamefont
  {A.}~\bibnamefont {Stepanov}},\ }\href@noop {} {\bibfield  {journal}
  {\bibinfo  {journal} {Journal of Materials Chemistry}\ }\textbf {\bibinfo
  {volume} {11}},\ \bibinfo {pages} {1152} (\bibinfo {year}
  {2001})}\BibitemShut {NoStop}%
\bibitem [{\citenamefont {Pregelj}\ \emph {et~al.}(2012)\citenamefont
  {Pregelj}, \citenamefont {Zaharko}, \citenamefont {G\"{u}nther},
  \citenamefont {Loidl}, \citenamefont {Tsurkan},\ and\ \citenamefont
  {Guerrero}}]{Pregelj2012}%
  \BibitemOpen
  \bibfield  {author} {\bibinfo {author} {\bibfnamefont {M.}~\bibnamefont
  {Pregelj}}, \bibinfo {author} {\bibfnamefont {O.}~\bibnamefont {Zaharko}},
  \bibinfo {author} {\bibfnamefont {A.}~\bibnamefont {G\"{u}nther}}, \bibinfo
  {author} {\bibfnamefont {A.}~\bibnamefont {Loidl}}, \bibinfo {author}
  {\bibfnamefont {V.}~\bibnamefont {Tsurkan}}, \ and\ \bibinfo {author}
  {\bibfnamefont {S.}~\bibnamefont {Guerrero}},\ }\href@noop {} {\bibfield
  {journal} {\bibinfo  {journal} {Physical Review B}\ }\textbf {\bibinfo
  {volume} {86}},\ \bibinfo {pages} {144409} (\bibinfo {year}
  {2012})}\BibitemShut {NoStop}%
\bibitem [{\citenamefont {Rousochatzakis}\ \emph {et~al.}(2015)\citenamefont
  {Rousochatzakis}, \citenamefont {Richter}, \citenamefont {Zinke},\ and\
  \citenamefont {Tsirlin}}]{Rousochatzakis2015}%
  \BibitemOpen
  \bibfield  {author} {\bibinfo {author} {\bibfnamefont {I.}~\bibnamefont
  {Rousochatzakis}}, \bibinfo {author} {\bibfnamefont {J.}~\bibnamefont
  {Richter}}, \bibinfo {author} {\bibfnamefont {R.}~\bibnamefont {Zinke}}, \
  and\ \bibinfo {author} {\bibfnamefont {A.~A.}\ \bibnamefont {Tsirlin}},\
  }\href@noop {} {\bibfield  {journal} {\bibinfo  {journal} {Physical Review
  B}\ }\textbf {\bibinfo {volume} {91}},\ \bibinfo {pages} {024416} (\bibinfo
  {year} {2015})}\BibitemShut {NoStop}%
\bibitem [{\citenamefont {Zakharov}\ \emph {et~al.}(2014)\citenamefont
  {Zakharov}, \citenamefont {Zvereva}, \citenamefont {Berdonosov},
  \citenamefont {Kuznetsova}, \citenamefont {Dolgikh}, \citenamefont {Clark},
  \citenamefont {Black}, \citenamefont {Lightfoot}, \citenamefont {Kockelmann},
  \citenamefont {Pchelkina}, \citenamefont {Streltsov}, \citenamefont
  {Volkova},\ and\ \citenamefont {Vasiliev}}]{Zakharov2014}%
  \BibitemOpen
  \bibfield  {author} {\bibinfo {author} {\bibfnamefont {K.~V.}\ \bibnamefont
  {Zakharov}}, \bibinfo {author} {\bibfnamefont {E.~A.}\ \bibnamefont
  {Zvereva}}, \bibinfo {author} {\bibfnamefont {P.~S.}\ \bibnamefont
  {Berdonosov}}, \bibinfo {author} {\bibfnamefont {E.~S.}\ \bibnamefont
  {Kuznetsova}}, \bibinfo {author} {\bibfnamefont {V.~A.}\ \bibnamefont
  {Dolgikh}}, \bibinfo {author} {\bibfnamefont {L.}~\bibnamefont {Clark}},
  \bibinfo {author} {\bibfnamefont {C.}~\bibnamefont {Black}}, \bibinfo
  {author} {\bibfnamefont {P.}~\bibnamefont {Lightfoot}}, \bibinfo {author}
  {\bibfnamefont {W.}~\bibnamefont {Kockelmann}}, \bibinfo {author}
  {\bibfnamefont {Z.~V.}\ \bibnamefont {Pchelkina}}, \bibinfo {author}
  {\bibfnamefont {S.~V.}\ \bibnamefont {Streltsov}}, \bibinfo {author}
  {\bibfnamefont {O.~S.}\ \bibnamefont {Volkova}}, \ and\ \bibinfo {author}
  {\bibfnamefont {A.~N.}\ \bibnamefont {Vasiliev}},\ }\href@noop {} {\bibfield
  {journal} {\bibinfo  {journal} {Physical Review B}\ }\textbf {\bibinfo
  {volume} {90}},\ \bibinfo {pages} {214417} (\bibinfo {year}
  {2014})}\BibitemShut {NoStop}%
\bibitem [{\citenamefont {Miller}\ \emph {et~al.}(2012)\citenamefont {Miller},
  \citenamefont {Stephens}, \citenamefont {Martin}, \citenamefont {Constable},
  \citenamefont {Lewis}, \citenamefont {Berger}, \citenamefont {Carr},\ and\
  \citenamefont {Tanner}}]{Miller2012}%
  \BibitemOpen
  \bibfield  {author} {\bibinfo {author} {\bibfnamefont {K.~H.}\ \bibnamefont
  {Miller}}, \bibinfo {author} {\bibfnamefont {P.~W.}\ \bibnamefont
  {Stephens}}, \bibinfo {author} {\bibfnamefont {C.}~\bibnamefont {Martin}},
  \bibinfo {author} {\bibfnamefont {E.}~\bibnamefont {Constable}}, \bibinfo
  {author} {\bibfnamefont {R.~A.}\ \bibnamefont {Lewis}}, \bibinfo {author}
  {\bibfnamefont {H.}~\bibnamefont {Berger}}, \bibinfo {author} {\bibfnamefont
  {G.~L.}\ \bibnamefont {Carr}}, \ and\ \bibinfo {author} {\bibfnamefont
  {D.~B.}\ \bibnamefont {Tanner}},\ }\href@noop {} {\bibfield  {journal}
  {\bibinfo  {journal} {Physical Review B}\ }\textbf {\bibinfo {volume} {86}},\
  \bibinfo {pages} {174104} (\bibinfo {year} {2012})}\BibitemShut {NoStop}%
\bibitem [{\citenamefont {Gnezdilov}\ \emph {et~al.}(2016)\citenamefont
  {Gnezdilov}, \citenamefont {Pashkevich}, \citenamefont {Kurnosov},
  \citenamefont {Lemmens}, \citenamefont {Kuznetsova}, \citenamefont
  {Berdonosov}, \citenamefont {Dolgikh}, \citenamefont {Zakharov},\ and\
  \citenamefont {Vasiliev}}]{Gnezdilov2016_arXiv}%
  \BibitemOpen
  \bibfield  {author} {\bibinfo {author} {\bibfnamefont {V.}~\bibnamefont
  {Gnezdilov}}, \bibinfo {author} {\bibfnamefont {Y.}~\bibnamefont
  {Pashkevich}}, \bibinfo {author} {\bibfnamefont {V.}~\bibnamefont
  {Kurnosov}}, \bibinfo {author} {\bibfnamefont {P.}~\bibnamefont {Lemmens}},
  \bibinfo {author} {\bibfnamefont {E.}~\bibnamefont {Kuznetsova}}, \bibinfo
  {author} {\bibfnamefont {P.}~\bibnamefont {Berdonosov}}, \bibinfo {author}
  {\bibfnamefont {V.}~\bibnamefont {Dolgikh}}, \bibinfo {author} {\bibfnamefont
  {K.}~\bibnamefont {Zakharov}}, \ and\ \bibinfo {author} {\bibfnamefont
  {A.}~\bibnamefont {Vasiliev}},\ }\href@noop {} {\bibfield  {journal}
  {\bibinfo  {journal} {arXiv:1604.04249}\ } (\bibinfo {year}
  {2016})}\BibitemShut {NoStop}%
\bibitem [{\citenamefont {Momma}\ and\ \citenamefont
  {Izumi}(2011)}]{Vesta_2011}%
  \BibitemOpen
  \bibfield  {author} {\bibinfo {author} {\bibfnamefont {K.}~\bibnamefont
  {Momma}}\ and\ \bibinfo {author} {\bibfnamefont {F.}~\bibnamefont {Izumi}},\
  }\href {\doibase 10.1107/s0021889811038970} {\bibfield  {journal} {\bibinfo
  {journal} {Journal of Applied Crystallography}\ }\textbf {\bibinfo {volume}
  {44}},\ \bibinfo {pages} {1272} (\bibinfo {year} {2011})}\BibitemShut
  {NoStop}%
\bibitem [{\citenamefont {Prishchenko}\ \emph {et~al.}(2017)\citenamefont
  {Prishchenko}, \citenamefont {Tsirlin}, \citenamefont {Tsurkan},
  \citenamefont {Loidl}, \citenamefont {Jesche},\ and\ \citenamefont
  {Mazurenko}}]{Prishchenko2017}%
  \BibitemOpen
  \bibfield  {author} {\bibinfo {author} {\bibfnamefont {D.~A.}\ \bibnamefont
  {Prishchenko}}, \bibinfo {author} {\bibfnamefont {A.~A.}\ \bibnamefont
  {Tsirlin}}, \bibinfo {author} {\bibfnamefont {V.}~\bibnamefont {Tsurkan}},
  \bibinfo {author} {\bibfnamefont {A.}~\bibnamefont {Loidl}}, \bibinfo
  {author} {\bibfnamefont {A.}~\bibnamefont {Jesche}}, \ and\ \bibinfo {author}
  {\bibfnamefont {V.~G.}\ \bibnamefont {Mazurenko}},\ }\href@noop {} {\bibfield
   {journal} {\bibinfo  {journal} {Physical Review B}\ }\textbf {\bibinfo
  {volume} {95}},\ \bibinfo {pages} {064102} (\bibinfo {year}
  {2017})}\BibitemShut {NoStop}%
\bibitem [{\citenamefont {Wu}\ \emph {et~al.}(2017)\citenamefont {Wu},
  \citenamefont {Chandrasekhar}, \citenamefont {Yuan}, \citenamefont {Huang},
  \citenamefont {Lin}, \citenamefont {Berger},\ and\ \citenamefont
  {Yang}}]{Wu2017}%
  \BibitemOpen
  \bibfield  {author} {\bibinfo {author} {\bibfnamefont {H.~C.}\ \bibnamefont
  {Wu}}, \bibinfo {author} {\bibfnamefont {K.~D.}\ \bibnamefont
  {Chandrasekhar}}, \bibinfo {author} {\bibfnamefont {J.~K.}\ \bibnamefont
  {Yuan}}, \bibinfo {author} {\bibfnamefont {J.~R.}\ \bibnamefont {Huang}},
  \bibinfo {author} {\bibfnamefont {J.-Y.}\ \bibnamefont {Lin}}, \bibinfo
  {author} {\bibfnamefont {H.}~\bibnamefont {Berger}}, \ and\ \bibinfo {author}
  {\bibfnamefont {H.~D.}\ \bibnamefont {Yang}},\ }\href@noop {} {\bibfield
  {journal} {\bibinfo  {journal} {Physical Review B}\ }\textbf {\bibinfo
  {volume} {95}},\ \bibinfo {pages} {125121} (\bibinfo {year}
  {2017})}\BibitemShut {NoStop}%
\bibitem [{\citenamefont {Aroyo}\ \emph
  {et~al.}(2006{\natexlab{a}})\citenamefont {Aroyo}, \citenamefont {Kirov},
  \citenamefont {Capillas}, \citenamefont {Perez-Mato},\ and\ \citenamefont
  {Wondratschek}}]{Aroyo_2006}%
  \BibitemOpen
  \bibfield  {author} {\bibinfo {author} {\bibfnamefont {M.~I.}\ \bibnamefont
  {Aroyo}}, \bibinfo {author} {\bibfnamefont {A.}~\bibnamefont {Kirov}},
  \bibinfo {author} {\bibfnamefont {C.}~\bibnamefont {Capillas}}, \bibinfo
  {author} {\bibfnamefont {J.~M.}\ \bibnamefont {Perez-Mato}}, \ and\ \bibinfo
  {author} {\bibfnamefont {H.}~\bibnamefont {Wondratschek}},\ }\href@noop {}
  {\bibfield  {journal} {\bibinfo  {journal} {Acta Crystallographica Section A
  Foundations of Crystallography}\ }\textbf {\bibinfo {volume} {62}},\ \bibinfo
  {pages} {115} (\bibinfo {year} {2006}{\natexlab{a}})}\BibitemShut {NoStop}%
\bibitem [{\citenamefont {Aroyo}\ \emph
  {et~al.}(2006{\natexlab{b}})\citenamefont {Aroyo}, \citenamefont
  {Perez-Mato}, \citenamefont {Capillas}, \citenamefont {Kroumova},
  \citenamefont {Ivantchev}, \citenamefont {Madariaga}, \citenamefont {Kirov},\
  and\ \citenamefont {Wondratschek}}]{Aroyo_2006_2}%
  \BibitemOpen
  \bibfield  {author} {\bibinfo {author} {\bibfnamefont {M.~I.}\ \bibnamefont
  {Aroyo}}, \bibinfo {author} {\bibfnamefont {J.~M.}\ \bibnamefont
  {Perez-Mato}}, \bibinfo {author} {\bibfnamefont {C.}~\bibnamefont
  {Capillas}}, \bibinfo {author} {\bibfnamefont {E.}~\bibnamefont {Kroumova}},
  \bibinfo {author} {\bibfnamefont {S.}~\bibnamefont {Ivantchev}}, \bibinfo
  {author} {\bibfnamefont {G.}~\bibnamefont {Madariaga}}, \bibinfo {author}
  {\bibfnamefont {A.}~\bibnamefont {Kirov}}, \ and\ \bibinfo {author}
  {\bibfnamefont {H.}~\bibnamefont {Wondratschek}},\ }\href@noop {} {\bibfield
  {journal} {\bibinfo  {journal} {Zeitschrift f{\"u}r Kristallographie -
  Crystalline Materials}\ }\textbf {\bibinfo {volume} {221}},\ \bibinfo {pages}
  {15} (\bibinfo {year} {2006}{\natexlab{b}})}\BibitemShut {NoStop}%
\bibitem [{\citenamefont {Aroyo}\ \emph {et~al.}(2011)\citenamefont {Aroyo},
  \citenamefont {Perez-Mato}, \citenamefont {Orobengoa}, \citenamefont {Tasci},
  \citenamefont {de~la Flor},\ and\ \citenamefont {Kirov}}]{Bilbao_2011}%
  \BibitemOpen
  \bibfield  {author} {\bibinfo {author} {\bibfnamefont {M.~I.}\ \bibnamefont
  {Aroyo}}, \bibinfo {author} {\bibfnamefont {J.~M.}\ \bibnamefont
  {Perez-Mato}}, \bibinfo {author} {\bibfnamefont {D.}~\bibnamefont
  {Orobengoa}}, \bibinfo {author} {\bibfnamefont {E.}~\bibnamefont {Tasci}},
  \bibinfo {author} {\bibfnamefont {G.}~\bibnamefont {de~la Flor}}, \ and\
  \bibinfo {author} {\bibfnamefont {A.}~\bibnamefont {Kirov}},\ }\href@noop {}
  {\bibfield  {journal} {\bibinfo  {journal} {Bulgarian Chemical
  Communications}\ }\textbf {\bibinfo {volume} {43}},\ \bibinfo {pages} {183}
  (\bibinfo {year} {2011})}\BibitemShut {NoStop}%
\bibitem [{\citenamefont {Roisnel}\ and\ \citenamefont
  {Rodr\'{\i}quez-Carvajal}(2001)}]{Roisnel_2001}%
  \BibitemOpen
  \bibfield  {author} {\bibinfo {author} {\bibfnamefont {T.}~\bibnamefont
  {Roisnel}}\ and\ \bibinfo {author} {\bibfnamefont {J.}~\bibnamefont
  {Rodr\'{\i}quez-Carvajal}},\ }\href {\doibase
  10.4028/www.scientific.net/msf.378-381.118} {\bibfield  {journal} {\bibinfo
  {journal} {Materials Science Forum}\ }\textbf {\bibinfo {volume} {378-381}},\
  \bibinfo {pages} {118} (\bibinfo {year} {2001})}\BibitemShut {NoStop}%
\bibitem [{\citenamefont
  {Rodr\'{\i}guez-Carvajal}(1993)}]{Rodriguez_Carvajal_1993}%
  \BibitemOpen
  \bibfield  {author} {\bibinfo {author} {\bibfnamefont {J.}~\bibnamefont
  {Rodr\'{\i}guez-Carvajal}},\ }\href {\doibase 10.1016/0921-4526(93)90108-i}
  {\bibfield  {journal} {\bibinfo  {journal} {Physica B: Condensed Matter}\
  }\textbf {\bibinfo {volume} {192}},\ \bibinfo {pages} {55} (\bibinfo {year}
  {1993})}\BibitemShut {NoStop}%
\bibitem [{\citenamefont {Petit}\ and\ \citenamefont {Damay}(2016)}]{SpinWave}%
  \BibitemOpen
  \bibfield  {author} {\bibinfo {author} {\bibfnamefont {S.}~\bibnamefont
  {Petit}}\ and\ \bibinfo {author} {\bibfnamefont {F.}~\bibnamefont {Damay}},\
  }\href@noop {} {\bibfield  {journal} {\bibinfo  {journal} {Neutron News}\
  }\textbf {\bibinfo {volume} {27}},\ \bibinfo {pages} {27} (\bibinfo {year}
  {2016})}\BibitemShut {NoStop}%
\bibitem [{\citenamefont {Tol\'{e}dano}\ and\ \citenamefont
  {Guennou}(2016)}]{Toledano_2016}%
  \BibitemOpen
  \bibfield  {author} {\bibinfo {author} {\bibfnamefont {P.}~\bibnamefont
  {Tol\'{e}dano}}\ and\ \bibinfo {author} {\bibfnamefont {M.}~\bibnamefont
  {Guennou}},\ }\href {http://dx.doi.org/10.1103/PhysRevB.94.014107} {\bibfield
   {journal} {\bibinfo  {journal} {Physical Review B}\ }\textbf {\bibinfo
  {volume} {94}},\ \bibinfo {pages} {014107} (\bibinfo {year}
  {2016})}\BibitemShut {NoStop}%
\bibitem [{\citenamefont {Rabe}(2013)}]{Rabe_2013}%
  \BibitemOpen
  \bibfield  {author} {\bibinfo {author} {\bibfnamefont {K.~M.}\ \bibnamefont
  {Rabe}},\ }\href {\doibase 10.1002/9783527654864.ch7} {\bibfield  {journal}
  {\bibinfo  {journal} {Functional Metal Oxides}\ ,\ \bibinfo {pages} {221}}
  (\bibinfo {year} {2013})}\BibitemShut {NoStop}%
\bibitem [{\citenamefont {Devonshire}(1954)}]{Devonshire_1954}%
  \BibitemOpen
  \bibfield  {author} {\bibinfo {author} {\bibfnamefont {A.}~\bibnamefont
  {Devonshire}},\ }\href {\doibase 10.1080/00018735400101173} {\bibfield
  {journal} {\bibinfo  {journal} {Advances in Physics}\ }\textbf {\bibinfo
  {volume} {3}},\ \bibinfo {pages} {85} (\bibinfo {year} {1954})}\BibitemShut
  {NoStop}%
\bibitem [{\citenamefont {Kittel}(1951)}]{Kittel_1951}%
  \BibitemOpen
  \bibfield  {author} {\bibinfo {author} {\bibfnamefont {C.}~\bibnamefont
  {Kittel}},\ }\href {\doibase 10.1103/physrev.82.729} {\bibfield  {journal}
  {\bibinfo  {journal} {Physical Review}\ }\textbf {\bibinfo {volume} {82}},\
  \bibinfo {pages} {729} (\bibinfo {year} {1951})}\BibitemShut {NoStop}%
\bibitem [{\citenamefont {Borovik-Romanov}\ and\ \citenamefont
  {Grimmer}(2006)}]{IntTbs_Crst_D_2013}%
  \BibitemOpen
  \bibfield  {author} {\bibinfo {author} {\bibfnamefont {A.~S.}\ \bibnamefont
  {Borovik-Romanov}}\ and\ \bibinfo {author} {\bibfnamefont {H.}~\bibnamefont
  {Grimmer}},\ }\href@noop {} {\bibfield  {journal} {\bibinfo  {journal}
  {International Tables for Crystallography, Vol. D (IUCr, Chester, UK),}\
  \bibinfo {pages} {Chap. 1.5, p. 138}} (\bibinfo {year} {2006})}\BibitemShut
  {NoStop}%
\bibitem [{\citenamefont {Gorbatsevich}\ \emph {et~al.}(1983)\citenamefont
  {Gorbatsevich}, \citenamefont {Kopaev},\ and\ \citenamefont
  {Tugushev}}]{Gorbatsevich_1983}%
  \BibitemOpen
  \bibfield  {author} {\bibinfo {author} {\bibfnamefont {A.~A.}\ \bibnamefont
  {Gorbatsevich}}, \bibinfo {author} {\bibfnamefont {Y.~V.}\ \bibnamefont
  {Kopaev}}, \ and\ \bibinfo {author} {\bibfnamefont {V.~V.}\ \bibnamefont
  {Tugushev}},\ }\href@noop {} {\bibfield  {journal} {\bibinfo  {journal}
  {Journal of Experimental and Theoretical Physics}\ }\textbf {\bibinfo
  {volume} {58}},\ \bibinfo {pages} {643} (\bibinfo {year} {1983})}\BibitemShut
  {NoStop}%
\end{thebibliography}

%

\end{document}